%%%%%%%%%%%%%%%%%%%%%%%%%%%%%%%%%%%%%%%%%%%%%%%%%%%%%%%%%%%%%%%%%%%%%
%
%
%                                      Andrea PASQUINUCCI, 1988
%              PANDA.TEX               S.I.S.S.A., Trieste, Italy
%                                      (Revised 1991, Princeton, USA)
%
%--------------------------------------------------------------------
%
%    These are TEX macros. They work with PLAIN TEX (the basis
%    version of TEX). The only problem can be with the double-page
%    format since it depends on the type of software and laserwriter
%    you use to print, so I cannot guarantee that the double-page
%    format will work properly. Double-page MUST be printed in
%    LANDSCAPE orientation. (You shouldn't have troubles with fonts;
%    if you do, please let me know.)
%
%--------------------------------------------------------------------
%
%                     INTERACTIVE SECTION
%
%--------------------------------------------------------------------
%
\def\standardrisposta{s }\def\reducedrisposta{r }
\def\mplarisposta{mpla }\def\zerorisposta{z }
\def\doublerisposta{d }\def\cartarisposta{e }\def\amsrisposta{y }
\newcount\ingrandimento \newcount\sinnota \newcount\dimnota
\newcount\unoduecol \newdimen\collhsize \newdimen\tothsize
\newdimen\fullhsize \newcount\controllorisposta \sinnota=1
\newskip\infralinea  \global\controllorisposta=0
\immediate\write16 { ********  Welcome to PANDA macros (Plain TeX,
AP, 1991) ******** }
%\immediate\write16 { You'll have to answer a few questions in
%lowercase.}
%\message{>  Do you want it in double-page (d), reduced (r)
%or standard format (s) ? }\read-1 to\risposta
%
%\message{>  Do you want it in USA A4 (u) or EUROPEAN A4
%(e) paper size ? }\read-1 to\srisposta
%
%\message{>  Do you have AMSFonts 2.0 (math) fonts (y/n) ? }
%\read-1 to\arisposta
%
%--------------------------------------------------------------------
%
%             END INTERACTIVE SECTION - PAGE FORMATTING
%
%--------------------------------------------------------------------
%       The following parameters define defaults to the interactive
%       session.  At the moment I have set EUROPEAN and MATH FONTS
%
\def\risposta{s } 
\def\srisposta{e } 
\def\arisposta{y }
\ifx\risposta\standardrisposta \ingrandimento=1200
\message {>> This will come out UNREDUCED << }
\dimnota=2 \unoduecol=1 \global\controllorisposta=1 \fi
\ifx\risposta\reducedrisposta \ingrandimento=1095 \dimnota=1
\unoduecol=1  \global\controllorisposta=1
\message {>> This will come out REDUCED << } \fi
\ifx\risposta\doublerisposta \ingrandimento=1000 \dimnota=2
\unoduecol=2

\message {>> You must print this in
LANDSCAPE orientation << } \global\controllorisposta=1 \fi
\ifx\risposta\mplarisposta \ingrandimento=1000 \dimnota=1
\message {>> Mod. Phys. Lett. A format << }
\unoduecol=1 \global\controllorisposta=1 \fi
\ifx\risposta\zerorisposta \ingrandimento=1000 \dimnota=2
\message {>> Zero Magnification format << }
\unoduecol=1 \global\controllorisposta=1 \fi
\ifnum\controllorisposta=0  \ingrandimento=1200
\message {>>> ERROR IN INPUT, I ASSUME STANDARD
UNREDUCED FORMAT <<< }  \dimnota=2 \unoduecol=1 \fi
\magnification=\ingrandimento
%
%--------------------------------------------------------------------
%
%                        PARAMETERS SETTING
%
%  You can modify these parameters at your will (and resposability)
%--------------------------------------------------------------------
%
\newdimen\eucolumnsize \newdimen\eudoublehsize \newdimen\eudoublevsize
\newdimen\uscolumnsize \newdimen\usdoublehsize \newdimen\usdoublevsize
\newdimen\eusinglehsize \newdimen\eusinglevsize \newdimen\ussinglehsize
\newskip\standardbaselineskip \newdimen\ussinglevsize
\newskip\reducedbaselineskip \newskip\doublebaselineskip
\eucolumnsize=12.0truecm    % column h-size for european doublepage
                            % (12.0treucm default)
\eudoublehsize=25.5truecm   % sheet h-size for european duoblepage
                            % (25.5treucm default)
\eudoublevsize=6.7truein    % sheet v-size for european doublepage
                            % (6.5treuin default  or 17truecm?)
\uscolumnsize=4.4truein     % column h-size for american doublepage
                            % (4.4treuin default)
\usdoublehsize=9.4truein    % sheet h-size for american duoblepage
                            % (9.4treuin default)
\usdoublevsize=6.8truein    % sheet v-size for american doublepage
                            % (6.8treuin default)
\eusinglehsize=6.5truein    % sheet h-size for european singlepage
                            % (6.5truein default)
\eusinglevsize=24truecm     % sheet v-size for european singlepage
                            % (24truecm default)
\ussinglehsize=6.5truein    % sheet h-size for american singlepage
                            % (6.5truein default)
\ussinglevsize=8.9truein    % sheet v-size for american singlepage
                            % (8.9truein default)
\standardbaselineskip=16pt plus.2pt  % baselineskip for standard
                                     % format (16pt default)
\reducedbaselineskip=14pt plus.2pt   % baselineskip for reduced
                                     % format (14pt default)
\doublebaselineskip=12pt plus.2pt    % baselineskip for doublepage
                                     % format (12pt default)
%
%  \Portoffset and \Landoffset define the horizontal and vertical
%  offsets respectively for portrait and landscape modes. Example:
%  \def\Portoffset{\voffset=.4truein\hoffset=.125truein}
%
\def\Portoffset{}
\def\Landoffset{\voffset=-.2truein}
\ifx\risposta\mplarisposta \def\Portoffset{\hoffset=1.8truecm} \fi
%
%  \Landspec defines the \special command that sets the printer
%  to landscape mode without need to specify it directly in the
%  TeX to postscript translator (the command is site dependent).
%  Example: \def\Landspec{\special{ps: landscape}}
%
\def\Landspec{}
\tolerance=10000
\parskip=0pt plus2pt  \leftskip=0pt \rightskip=0pt
%
%   Do not modify anything of what follows
%                       (unless you know what you are doing!)
%----------------------------------------------------------------------
%
\ifx\risposta\standardrisposta \infralinea=\standardbaselineskip \fi
\ifx\risposta\reducedrisposta  \infralinea=\reducedbaselineskip \fi
\ifx\risposta\doublerisposta   \infralinea=\doublebaselineskip \fi
\ifx\risposta\mplarisposta     \infralinea=13pt \fi
\ifx\risposta\zerorisposta     \infralinea=12pt plus.2pt\fi
\ifnum\controllorisposta=0    \infralinea=\standardbaselineskip \fi
\ifx\risposta\doublerisposta   \Landoffset \else \Portoffset \fi
\ifx\risposta\doublerisposta \ifx\srisposta\cartarisposta
\tothsize=\eudoublehsize \collhsize=\eucolumnsize
\vsize=\eudoublevsize  \else  \tothsize=\usdoublehsize
\collhsize=\uscolumnsize \vsize=\usdoublevsize \fi \else
\ifx\srisposta\cartarisposta \tothsize=\eusinglehsize
\vsize=\eusinglevsize \else  \tothsize=\ussinglehsize
\vsize=\ussinglevsize \fi \collhsize=4.4truein \fi
\ifx\risposta\mplarisposta \tothsize=5.0truein
\vsize=7.8truein \collhsize=4.4truein \fi
%
%--------------------------------------------------------------------
%
%                            FONTS
%
%--------------------------------------------------------------------
%
\newcount\contaeuler \newcount\contacyrill \newcount\contaams
\font\ninerm=cmr9  \font\eightrm=cmr8  \font\sixrm=cmr6
\font\ninei=cmmi9  \font\eighti=cmmi8  \font\sixi=cmmi6
\font\ninesy=cmsy9  \font\eightsy=cmsy8  \font\sixsy=cmsy6
\font\ninebf=cmbx9  \font\eightbf=cmbx8  \font\sixbf=cmbx6
\font\ninett=cmtt9  \font\eighttt=cmtt8  \font\nineit=cmti9
\font\eightit=cmti8 \font\ninesl=cmsl9  \font\eightsl=cmsl8
\skewchar\ninei='177 \skewchar\eighti='177 \skewchar\sixi='177
\skewchar\ninesy='60 \skewchar\eightsy='60 \skewchar\sixsy='60
\hyphenchar\ninett=-1 \hyphenchar\eighttt=-1 \hyphenchar\tentt=-1
\def\bfmath{\cmmib}                 % math italic bold \bfmath
\font\tencmmib=cmmib10  \newfam\cmmibfam  \skewchar\tencmmib='177
                  % math bold (cal) symbols
\font\tencmbsy=cmbsy10  \newfam\cmbsyfam  \skewchar\tencmbsy='60
\def\scaps{\cmcsc}                 % small caps (uppercase)
\font\tencmcsc=cmcsc10  \newfam\cmcscfam
\ifnum\ingrandimento=1095

\font\capsone=cmcsc10 at 10.95pt \font\capstwo=cmcsc10 at 13.145pt

\else

\font\capsone=cmcsc10 at 12pt \font\capstwo=cmcsc10 at 14.4pt
\fi

\def\ttaarr{\bf}		% chapter titles' font
\def\ppaarr{\sl}		% section titles' font

%
     % inch-high caps (enormous)
%
%   AMS fonts (this works only if you have at least the 2.0
%              version of AMSFonts, otherwise say no)
%
\newfam\eufmfam \newfam\msamfam \newfam\msbmfam \newfam\eufbfam
\def\Loadeulerfonts{\global\contaeuler=1 \ifx\arisposta\amsrisposta
\font\teneufm=eufm10              %  \eufm   Gothic (or Euler)
\font\eighteufm=eufm8 \font\nineeufm=eufm9 \font\sixeufm=eufm6
\font\seveneufm=eufm7  \font\fiveeufm=eufm5
\font\teneufb=eufb10              %  \eufb   Bold Gothic (or Euler)
\font\eighteufb=eufb8 \font\nineeufb=eufb9 \font\sixeufb=eufb6
\font\seveneufb=eufb7  \font\fiveeufb=eufb5
\font\teneurm=eurm10              %  \eurm   Roman Gothic (or Euler)
\font\eighteurm=eurm8 \font\nineeurm=eurm9
\font\teneurb=eurb10              %  \eurb   Roman Bold Gothic
\font\eighteurb=eurb8 \font\nineeurb=eurb9
\font\teneusm=eusm10              %  \eusm   Slanted Capital Gothic
\font\eighteusm=eusm8 \font\nineeusm=eusm9
\font\teneusb=eusb10              %\eusb Slanted Capital Bold Gothic
\font\eighteusb=eusb8 \font\nineeusb=eusb9
\else \def\eufm{\tt} \def\eufb{\tt} \def\eurm{\tt} \def\eurb{\tt}
\def\eusm{\tt} \def\eusb{\tt}    \fi}
\def\loadeuler{\Loadeulerfonts\tenpoint}
\def\loadamsmath{\global\contaams=1 \ifx\arisposta\amsrisposta
\font\tenmsam=msam10 \font\ninemsam=msam9 \font\eightmsam=msam8
\font\sevenmsam=msam7 \font\sixmsam=msam6 \font\fivemsam=msam5
\font\tenmsbm=msbm10 \font\ninemsbm=msbm9 \font\eightmsbm=msbm8
\font\sevenmsbm=msbm7 \font\sixmsbm=msbm6 \font\fivemsbm=msbm5
\else \def\msbm{\bf} \fi \def\Bbb{\msbm} \def\symbl{\msam} \tenpoint}
\def\loadcyrill{\global\contacyrill=1 \ifx\arisposta\amsrisposta
\font\tenwncyr=wncyr10 \font\ninewncyr=wncyr9 \font\eightwncyr=wncyr8
\font\tenwncyb=wncyr10 \font\ninewncyb=wncyr9 \font\eightwncyb=wncyr8
\font\tenwncyi=wncyr10 \font\ninewncyi=wncyr9 \font\eightwncyi=wncyr8
\else \def\cyrill{\sl} \def\cyrilb{\sl} \def\cyrili{\sl} \fi\tenpoint}
\ifx\arisposta\amsrisposta
\font\sevenex=cmex7               %  reduced math symbols
\font\eightex=cmex8  \font\nineex=cmex9
\font\ninecmmib=cmmib9   \font\eightcmmib=cmmib8
\font\sevencmmib=cmmib7 \font\sixcmmib=cmmib6
\font\fivecmmib=cmmib5   \skewchar\ninecmmib='177
\skewchar\eightcmmib='177  \skewchar\sevencmmib='177
\skewchar\sixcmmib='177   \skewchar\fivecmmib='177
\font\ninecmbsy=cmbsy9    \font\eightcmbsy=cmbsy8
\font\sevencmbsy=cmbsy7  \font\sixcmbsy=cmbsy6
\font\fivecmbsy=cmbsy5   \skewchar\ninecmbsy='60
\skewchar\eightcmbsy='60  \skewchar\sevencmbsy='60
\skewchar\sixcmbsy='60    \skewchar\fivecmbsy='60
\font\ninecmcsc=cmcsc9    \font\eightcmcsc=cmcsc8     \else
\def\cmmib{\fam\cmmibfam\tencmmib}\textfont\cmmibfam=\tencmmib
\scriptfont\cmmibfam=\tencmmib \scriptscriptfont\cmmibfam=\tencmmib
\def\cmbsy{\fam\cmbsyfam\tencmbsy} \textfont\cmbsyfam=\tencmbsy
\scriptfont\cmbsyfam=\tencmbsy \scriptscriptfont\cmbsyfam=\tencmbsy
\scriptfont\cmcscfam=\tencmcsc \scriptscriptfont\cmcscfam=\tencmcsc
\def\cmcsc{\fam\cmcscfam\tencmcsc} \textfont\cmcscfam=\tencmcsc \fi
\catcode`@=11
\newskip\ttglue
\gdef\tenpoint{\def\rm{\fam0\tenrm}
  \textfont0=\tenrm \scriptfont0=\sevenrm \scriptscriptfont0=\fiverm
  \textfont1=\teni \scriptfont1=\seveni \scriptscriptfont1=\fivei
  \textfont2=\tensy \scriptfont2=\sevensy \scriptscriptfont2=\fivesy
  \textfont3=\tenex \scriptfont3=\tenex \scriptscriptfont3=\tenex
  \def\mcal{\fam2 \tensy}  \def\mmit{\fam1 \teni}
  \textfont\itfam=\tenit \def\it{\fam\itfam\tenit}
  \textfont\slfam=\tensl \def\sl{\fam\slfam\tensl}
  \textfont\ttfam=\tentt \scriptfont\ttfam=\eighttt
  \scriptscriptfont\ttfam=\eighttt  \def\tt{\fam\ttfam\tentt}
  \textfont\bffam=\tenbf \scriptfont\bffam=\sevenbf
  \scriptscriptfont\bffam=\fivebf \def\bf{\fam\bffam\tenbf}
     \ifx\arisposta\amsrisposta    \ifnum\contaeuler=1
  \textfont\eufmfam=\teneufm \scriptfont\eufmfam=\seveneufm
  \scriptscriptfont\eufmfam=\fiveeufm \def\eufm{\fam\eufmfam\teneufm}
  \textfont\eufbfam=\teneufb \scriptfont\eufbfam=\seveneufb
  \scriptscriptfont\eufbfam=\fiveeufb \def\eufb{\fam\eufbfam\teneufb}
  \def\eurm{\teneurm} \def\eurb{\teneurb} \def\eusm{\teneusm}
  \def\eusb{\teneusb}    \fi    \ifnum\contaams=1
  \textfont\msamfam=\tenmsam \scriptfont\msamfam=\sevenmsam
  \scriptscriptfont\msamfam=\fivemsam \def\msam{\fam\msamfam\tenmsam}
  \textfont\msbmfam=\tenmsbm \scriptfont\msbmfam=\sevenmsbm
  \scriptscriptfont\msbmfam=\fivemsbm \def\msbm{\fam\msbmfam\tenmsbm}
     \fi      \ifnum\contacyrill=1     \def\cyrill{\tenwncyr}
  \def\cyrilb{\tenwncyb}  \def\cyrili{\tenwncyi}         \fi
  \textfont3=\tenex \scriptfont3=\sevenex \scriptscriptfont3=\sevenex
  \def\cmmib{\fam\cmmibfam\tencmmib} \scriptfont\cmmibfam=\sevencmmib
  \textfont\cmmibfam=\tencmmib  \scriptscriptfont\cmmibfam=\fivecmmib
  \def\cmbsy{\fam\cmbsyfam\tencmbsy} \scriptfont\cmbsyfam=\sevencmbsy
  \textfont\cmbsyfam=\tencmbsy  \scriptscriptfont\cmbsyfam=\fivecmbsy
  \def\cmcsc{\fam\cmcscfam\tencmcsc} \scriptfont\cmcscfam=\eightcmcsc
  \textfont\cmcscfam=\tencmcsc \scriptscriptfont\cmcscfam=\eightcmcsc
     \fi            \tt \ttglue=.5em plus.25em minus.15em
  \normalbaselineskip=12pt
  \setbox\strutbox=\hbox{\vrule height8.5pt depth3.5pt width0pt}
  \let\sc=\eightrm \let\big=\tenbig   \normalbaselines
  \baselineskip=\infralinea  \rm}
\gdef\ninepoint{\def\rm{\fam0\ninerm}
  \textfont0=\ninerm \scriptfont0=\sixrm \scriptscriptfont0=\fiverm
  \textfont1=\ninei \scriptfont1=\sixi \scriptscriptfont1=\fivei
  \textfont2=\ninesy \scriptfont2=\sixsy \scriptscriptfont2=\fivesy
  \textfont3=\tenex \scriptfont3=\tenex \scriptscriptfont3=\tenex
  \def\mcal{\fam2 \ninesy}  \def\mmit{\fam1 \ninei}
  \textfont\itfam=\nineit \def\it{\fam\itfam\nineit}
  \textfont\slfam=\ninesl \def\sl{\fam\slfam\ninesl}
  \textfont\ttfam=\ninett \scriptfont\ttfam=\eighttt
  \scriptscriptfont\ttfam=\eighttt \def\tt{\fam\ttfam\ninett}
  \textfont\bffam=\ninebf \scriptfont\bffam=\sixbf
  \scriptscriptfont\bffam=\fivebf \def\bf{\fam\bffam\ninebf}
     \ifx\arisposta\amsrisposta  \ifnum\contaeuler=1
  \textfont\eufmfam=\nineeufm \scriptfont\eufmfam=\sixeufm
  \scriptscriptfont\eufmfam=\fiveeufm \def\eufm{\fam\eufmfam\nineeufm}
  \textfont\eufbfam=\nineeufb \scriptfont\eufbfam=\sixeufb
  \scriptscriptfont\eufbfam=\fiveeufb \def\eufb{\fam\eufbfam\nineeufb}
  \def\eurm{\nineeurm} \def\eurb{\nineeurb} \def\eusm{\nineeusm}
  \def\eusb{\nineeusb}     \fi   \ifnum\contaams=1
  \textfont\msamfam=\ninemsam \scriptfont\msamfam=\sixmsam
  \scriptscriptfont\msamfam=\fivemsam \def\msam{\fam\msamfam\ninemsam}
  \textfont\msbmfam=\ninemsbm \scriptfont\msbmfam=\sixmsbm
  \scriptscriptfont\msbmfam=\fivemsbm \def\msbm{\fam\msbmfam\ninemsbm}
     \fi       \ifnum\contacyrill=1     \def\cyrill{\ninewncyr}
  \def\cyrilb{\ninewncyb}  \def\cyrili{\ninewncyi}         \fi
  \textfont3=\nineex \scriptfont3=\sevenex \scriptscriptfont3=\sevenex
  \def\cmmib{\fam\cmmibfam\ninecmmib}  \textfont\cmmibfam=\ninecmmib
  \scriptfont\cmmibfam=\sixcmmib \scriptscriptfont\cmmibfam=\fivecmmib
  \def\cmbsy{\fam\cmbsyfam\ninecmbsy}  \textfont\cmbsyfam=\ninecmbsy
  \scriptfont\cmbsyfam=\sixcmbsy \scriptscriptfont\cmbsyfam=\fivecmbsy
  \def\cmcsc{\fam\cmcscfam\ninecmcsc} \scriptfont\cmcscfam=\eightcmcsc
  \textfont\cmcscfam=\ninecmcsc \scriptscriptfont\cmcscfam=\eightcmcsc
     \fi            \tt \ttglue=.5em plus.25em minus.15em
  \normalbaselineskip=11pt
  \setbox\strutbox=\hbox{\vrule height8pt depth3pt width0pt}
  \let\sc=\sevenrm \let\big=\ninebig \normalbaselines\rm}
\gdef\eightpoint{\def\rm{\fam0\eightrm}
  \textfont0=\eightrm \scriptfont0=\sixrm \scriptscriptfont0=\fiverm
  \textfont1=\eighti \scriptfont1=\sixi \scriptscriptfont1=\fivei
  \textfont2=\eightsy \scriptfont2=\sixsy \scriptscriptfont2=\fivesy
  \textfont3=\tenex \scriptfont3=\tenex \scriptscriptfont3=\tenex
  \def\mcal{\fam2 \eightsy}  \def\mmit{\fam1 \eighti}
  \textfont\itfam=\eightit \def\it{\fam\itfam\eightit}
  \textfont\slfam=\eightsl \def\sl{\fam\slfam\eightsl}
  \textfont\ttfam=\eighttt \scriptfont\ttfam=\eighttt
  \scriptscriptfont\ttfam=\eighttt \def\tt{\fam\ttfam\eighttt}
  \textfont\bffam=\eightbf \scriptfont\bffam=\sixbf
  \scriptscriptfont\bffam=\fivebf \def\bf{\fam\bffam\eightbf}
     \ifx\arisposta\amsrisposta   \ifnum\contaeuler=1
  \textfont\eufmfam=\eighteufm \scriptfont\eufmfam=\sixeufm
  \scriptscriptfont\eufmfam=\fiveeufm \def\eufm{\fam\eufmfam\eighteufm}
  \textfont\eufbfam=\eighteufb \scriptfont\eufbfam=\sixeufb
  \scriptscriptfont\eufbfam=\fiveeufb \def\eufb{\fam\eufbfam\eighteufb}
  \def\eurm{\eighteurm} \def\eurb{\eighteurb} \def\eusm{\eighteusm}
  \def\eusb{\eighteusb}       \fi    \ifnum\contaams=1
  \textfont\msamfam=\eightmsam \scriptfont\msamfam=\sixmsam
  \scriptscriptfont\msamfam=\fivemsam \def\msam{\fam\msamfam\eightmsam}
  \textfont\msbmfam=\eightmsbm \scriptfont\msbmfam=\sixmsbm
  \scriptscriptfont\msbmfam=\fivemsbm \def\msbm{\fam\msbmfam\eightmsbm}
     \fi       \ifnum\contacyrill=1     \def\cyrill{\eightwncyr}
  \def\cyrilb{\eightwncyb}  \def\cyrili{\eightwncyi}         \fi
  \textfont3=\eightex \scriptfont3=\sevenex \scriptscriptfont3=\sevenex
  \def\cmmib{\fam\cmmibfam\eightcmmib}  \textfont\cmmibfam=\eightcmmib
  \scriptfont\cmmibfam=\sixcmmib \scriptscriptfont\cmmibfam=\fivecmmib
  \def\cmbsy{\fam\cmbsyfam\eightcmbsy}  \textfont\cmbsyfam=\eightcmbsy
  \scriptfont\cmbsyfam=\sixcmbsy \scriptscriptfont\cmbsyfam=\fivecmbsy
  \def\cmcsc{\fam\cmcscfam\eightcmcsc} \scriptfont\cmcscfam=\eightcmcsc
  \textfont\cmcscfam=\eightcmcsc \scriptscriptfont\cmcscfam=\eightcmcsc
     \fi             \tt \ttglue=.5em plus.25em minus.15em
  \normalbaselineskip=9pt
  \setbox\strutbox=\hbox{\vrule height7pt depth2pt width0pt}
  \let\sc=\sixrm \let\big=\eightbig \normalbaselines\rm }
\gdef\tenbig#1{{\hbox{$\left#1\vbox to8.5pt{}\right.\n@space$}}}
\gdef\ninebig#1{{\hbox{$\textfont0=\tenrm\textfont2=\tensy
   \left#1\vbox to7.25pt{}\right.\n@space$}}}
\gdef\eightbig#1{{\hbox{$\textfont0=\ninerm\textfont2=\ninesy
   \left#1\vbox to6.5pt{}\right.\n@space$}}}
 %for 10-pt math in 9-pt territory
\def\alternativefont#1#2{\ifx\arisposta\amsrisposta \relax \else
\xdef#1{#2} \fi}
\global\contaeuler=0 \global\contacyrill=0 \global\contaams=0
%
%--------------------------------------------------------------------
%
%                            MACROS
%
%--------------------------------------------------------------------
%
\newbox\fotlinebb \newbox\hedlinebb \newbox\leftcolumn
\gdef\makeheadline{\vbox to 0pt{\vskip-22.5pt
     \fullline{\vbox to8.5pt{}\the\headline}\vss}\nointerlineskip}
\gdef\makehedlinebb{\vbox to 0pt{\vskip-22.5pt
     \fullline{\vbox to8.5pt{}\copy\hedlinebb\hfil
     \line{\hfill\the\headline\hfill}}\vss} \nointerlineskip}
\gdef\makefootline{\baselineskip=24pt \fullline{\the\footline}}
\gdef\makefotlinebb{\baselineskip=24pt
    \fullline{\copy\fotlinebb\hfil\line{\hfill\the\footline\hfill}}}
\gdef\doubleformat{\shipout\vbox{\Landspec\makehedlinebb
     \fullline{\box\leftcolumn\hfil\columnbox}\makefotlinebb}
     \advancepageno}
\gdef\columnbox{\leftline{\pagebody}}
\gdef\line#1{\hbox to\hsize{\hskip\leftskip#1\hskip\rightskip}}
\gdef\fullline#1{\hbox to\fullhsize{\hskip\leftskip{#1}%
\hskip\rightskip}}
\gdef\footnote#1{\let\@sf=\empty
         \ifhmode\edef\#sf{\spacefactor=\the\spacefactor}\/\fi
         #1\@sf\vfootnote{#1}}
\gdef\vfootnote#1{\insert\footins\bgroup
         \ifnum\dimnota=1  \eightpoint\fi
         \ifnum\dimnota=2  \ninepoint\fi
         \ifnum\dimnota=0  \tenpoint\fi
         \interlinepenalty=\interfootnotelinepenalty
         \splittopskip=\ht\strutbox
         \splitmaxdepth=\dp\strutbox \floatingpenalty=20000
         \leftskip=\oldssposta \rightskip=\olddsposta
         \spaceskip=0pt \xspaceskip=0pt
         \ifnum\sinnota=0   \textindent{#1}\fi
         \ifnum\sinnota=1   \item{#1}\fi
         \footstrut\futurelet\next\fo@t}
\gdef\fo@t{\ifcat\bgroup\noexpand\next \let\next\f@@t
             \else\let\next\f@t\fi \next}
\gdef\f@@t{\bgroup\aftergroup\@foot\let\next}
\gdef\f@t#1{#1\@foot} \gdef\@foot{\strut\egroup}
\gdef\footstrut{\vbox to\splittopskip{}}
\skip\footins=\bigskipamount
\count\footins=1000  \dimen\footins=8in
\catcode`@=12
\tenpoint
\ifnum\unoduecol=1 \hsize=\tothsize   \fullhsize=\tothsize \fi
\ifnum\unoduecol=2 \hsize=\collhsize  \fullhsize=\tothsize \fi
\global\let\lrcol=L      \ifnum\unoduecol=1
\output{\plainoutput{\ifnum\tipbnota=2 \clearnmbnota\fi}} \fi
\ifnum\unoduecol=2 \output{\if L\lrcol
     \global\setbox\leftcolumn=\columnbox
     \global\setbox\fotlinebb=\line{\hfill\the\footline\hfill}
     \global\setbox\hedlinebb=\line{\hfill\the\headline\hfill}
     \advancepageno  \global\let\lrcol=R
     \else  \doubleformat \global\let\lrcol=L \fi
     \ifnum\outputpenalty>-20000 \else\dosupereject\fi
     \ifnum\tipbnota=2\clearnmbnota\fi }\fi
\def\ifdoublepage{\ifnum\unoduecol=2 }
\gdef\yespagenumbers{\footline={\hss\tenrm\folio\hss}}
\gdef\ciao{ \ifnum\fdefcontre=1 \endfdef\fi
     \par\vfill\supereject \ifnum\unoduecol=2
     \if R\lrcol  \headline={}\nopagenumbers\null\vfill\eject
     \fi\fi \end}

\newskip\olddsposta \newskip\oldssposta
\global\oldssposta=\leftskip \global\olddsposta=\rightskip

\def\filldots{\leaders\hbox to 1em{\hss.\hss}\hfill}
\def\inquadrb#1 {\vbox {\hrule  \hbox{\vrule \vbox {\vskip .2cm
    \hbox {\ #1\ } \vskip .2cm } \vrule  }  \hrule} }
 \def\newline{\hfil\break}
\def\jump{\vskip\baselineskip} \newskip\iinnffrr
\def\sjump{\iinnffrr=\baselineskip
          \divide\iinnffrr by 2 \vskip\iinnffrr}
\def\bjump{\vskip\baselineskip \vskip\baselineskip}
\newcount\nmbnota  \def\clearnmbnota{\global\nmbnota=0}
\newcount\tipbnota \def\letterfootnote{\global\tipbnota=1}

\def\note#1{\global\advance\nmbnota by 1 \ifnum\tipbnota=1
    \footnote{$^{\rm\nttlett}$}{#1} \else {\ifnum\tipbnota=2
    \footnote{$^{\nttsymb}$}{#1}
    \else\footnote{$^{\the\nmbnota}$}{#1}\fi}\fi}
\def\nttlett{\ifcase\nmbnota \or a\or b\or c\or d\or e\or f\or
g\or h\or i\or j\or k\or l\or m\or n\or o\or p\or q\or r\or
s\or t\or u\or v\or w\or y\or x\or z\fi}
\def\nttsymb{\ifcase\nmbnota \or\dag\or\sharp\or\ddag\or\star\or
\natural\or\flat\or\clubsuit\or\diamondsuit\or\heartsuit
\or\spadesuit\fi}   \clearnmbnota
\def\numberfootnote{\global\tipbnota=0} \numberfootnote
\def\setnote#1{\expandafter\xdef\csname#1\endcsname{
\ifnum\tipbnota=1 {\rm\nttlett} \else {\ifnum\tipbnota=2
{\nttsymb} \else \the\nmbnota\fi}\fi} }
\newcount\nbmfig  \def\clearnbmfig{\global\nbmfig=0}
\gdef\figure{\global\advance\nbmfig by 1
      {\rm fig. \the\nbmfig}}   \clearnbmfig
\def\setfig#1{\expandafter\xdef\csname#1\endcsname{fig. \the\nbmfig}}
 \def\endformula{\eqno\numero $$}
 \def\efr{\endformula}
\newcount\frmcount \def\clearfrmcount{\global\frmcount=0}
\def\numero{\global\advance\frmcount by 1   \ifnum\indappcount=0
  {\ifnum\cpcount <1 {\hbox{\rm (\the\frmcount )}}  \else
  {\hbox{\rm (\the\cpcount .\the\frmcount )}} \fi}  \else
  {\hbox{\rm (\applett .\the\frmcount )}} \fi}
\def\nameformula#1{\global\advance\frmcount by 1%
\ifnum\draftnum=0  {\ifnum\indappcount=0%
{\ifnum\cpcount<1\xdef\spzzttrra{(\the\frmcount )}%
\else\xdef\spzzttrra{(\the\cpcount .\the\frmcount )}\fi}%
\else\xdef\spzzttrra{(\applett .\the\frmcount )}\fi}%
\else\xdef\spzzttrra{(#1)}\fi%
\expandafter\xdef\csname#1\endcsname{\spzzttrra}
\eqno \hbox{\rm\spzzttrra} $$}
\def\nfr{\nameformula}    \def\numali{\numero}
\def\nameali#1{\global\advance\frmcount by 1%
\ifnum\draftnum=0  {\ifnum\indappcount=0%
{\ifnum\cpcount<1\xdef\spzzttrra{(\the\frmcount )}%
\else\xdef\spzzttrra{(\the\cpcount .\the\frmcount )}\fi}%
\else\xdef\spzzttrra{(\applett .\the\frmcount )}\fi}%
\else\xdef\spzzttrra{(#1)}\fi%
\expandafter\xdef\csname#1\endcsname{\spzzttrra}
  \hbox{\rm\spzzttrra} }      \clearfrmcount
\newcount\cpcount \def\clearcpcount{\global\cpcount=0}
\newcount\subcpcount \def\clearsubcpcount{\global\subcpcount=0}
\newcount\appcount \def\clearappcount{\global\appcount=0}
\newcount\indappcount \def\clearindappcount{\indappcount=0}
\newcount\sottoparcount 

\def\applett{\ifcase\appcount  \or {A}\or {B}\or {C}\or
{D}\or {E}\or {F}\or {G}\or {H}\or {I}\or {J}\or {K}\or {L}\or
{M}\or {N}\or {O}\or {P}\or {Q}\or {R}\or {S}\or {T}\or {U}\or
{V}\or {W}\or {X}\or {Y}\or {Z}\fi    \ifnum\appcount<0
\immediate\write16 {Panda ERROR - Appendix: counter "appcount"
out of range}\fi  \ifnum\appcount>26  \immediate\write16 {Panda
ERROR - Appendix: counter "appcount" out of range}\fi}
\clearappcount  \clearindappcount \newcount\connttrre
\def\clearconnttrre{\global\connttrre=0} \newcount\countref
\def\clearcountref{\global\countref=0} \clearcountref
\def\chapter#1{\global\advance\cpcount by 1 \clearfrmcount
                 \goodbreak\null\vbox{\jump\nobreak
                 \clearsubcpcount\clearindappcount
                 \itemitem{\ttaarr\the\cpcount .\qquad}{\ttaarr #1}
                 \par\nobreak\jump\sjump}\nobreak}
\def\section#1{\global\advance\subcpcount by 1 \goodbreak\null
               \vbox{\sjump\nobreak\ifnum\indappcount=0
                 {\ifnum\cpcount=0 {\itemitem{\ppaarr
               .\the\subcpcount\quad\enskip\ }{\ppaarr #1}\par} \else
                 {\itemitem{\ppaarr\the\cpcount .\the\subcpcount\quad
                  \enskip\ }{\ppaarr #1} \par}  \fi}
                \else{\itemitem{\ppaarr\applett .\the\subcpcount\quad
                 \enskip\ }{\ppaarr #1}\par}\fi\nobreak\jump}\nobreak}
\clearsubcpcount
\def\appendix#1{\global\advance\appcount by 1 \clearfrmcount
                  \goodbreak\null\vbox{\jump\nobreak
                  \global\advance\indappcount by 1 \clearsubcpcount
          \itemitem{ }{\hskip-40pt\ttaarr #1}
%                  \itemitem{\ttaarr App.\applett\ }{\ttaarr #1}
             \nobreak\jump\sjump}\nobreak}
\clearappcount \clearindappcount
\def\references{\goodbreak\null\vbox{\jump\nobreak
   \noindent{\ttaarr References} \nobreak\jump\sjump}\nobreak}
%   \itemitem{}{\ttaarr References} \nobreak\jump\sjump}\nobreak}

\clearcpcount\clearcountref

\def\setchap#1{\ifnum\indappcount=0{\ifnum\subcpcount=0%
\xdef\spzzttrra{\the\cpcount}%
\else\xdef\spzzttrra{\the\cpcount .\the\subcpcount}\fi}
\else{\ifnum\subcpcount=0 \xdef\spzzttrra{\applett}%
\else\xdef\spzzttrra{\applett .\the\subcpcount}\fi}\fi
\expandafter\xdef\csname#1\endcsname{\spzzttrra}}
\newcount\draftnum \newcount\ppora   \newcount\ppminuti
\global\ppora=\time   \global\ppminuti=\time
\global\divide\ppora by 60  \draftnum=\ppora
\multiply\draftnum by 60    \global\advance\ppminuti by -\draftnum
\def\droggi{\number\day /\number\month /\number\year\ \the\ppora
:\the\ppminuti}     \global\draftnum=0
\def\draftcomment#1{\ifnum\draftnum=0 \relax \else
{\ {\bf ***}\ #1\ {\bf ***}\ }\fi} 
%
%     Maximum number of references = 200
%     boxes 50 -> 250 reserved for references
%
\catcode`@=11
\gdef\Ref#1{\expandafter\ifx\csname @rrxx@#1\endcsname\relax%
{\global\advance\countref by 1    \ifnum\countref>200
\immediate\write16 {Panda ERROR - Ref: maximum number of references
exceeded}  \expandafter\xdef\csname @rrxx@#1\endcsname{0}\else
\expandafter\xdef\csname @rrxx@#1\endcsname{\the\countref}\fi}\fi
\ifnum\draftnum=0 \csname @rrxx@#1\endcsname \else#1\fi}
\gdef\beginref{\ifnum\draftnum=0  \gdef\Rref{\fairef}
\gdef\endref{\scriviref} \else\relax\fi
\ifx\risposta\mplarisposta \ninepoint \fi
\parskip 2pt plus.2pt \baselineskip=12pt}
\def\Reflab#1{[#1]} \gdef\Rref#1#2{\item{\Reflab{#1}}{#2}}
\gdef\endref{\relax}  \newcount\conttemp
\gdef\fairef#1#2{\expandafter\ifx\csname @rrxx@#1\endcsname\relax
{\global\conttemp=0 \immediate\write16 {Panda ERROR - Ref: reference
[#1] undefined}} \else
{\global\conttemp=\csname @rrxx@#1\endcsname } \fi
\global\advance\conttemp by 50  \global\setbox\conttemp=\hbox{#2} }
\gdef\scriviref{\clearconnttrre\conttemp=50
\loop\ifnum\connttrre<\countref \advance\conttemp by 1
\advance\connttrre by 1
\item{\Reflab{\the\connttrre}}{\unhcopy\conttemp} \repeat}
\clearcountref \clearconnttrre
\catcode`@=12
\ifx\risposta\mplarisposta \def\Reflab#1{#1.} \letterfootnote \fi

\def\slashchar#1{\setbox0=\hbox{$#1$} \dimen0=\wd0
     \setbox1=\hbox{/} \dimen1=\wd1 \ifdim\dimen0>\dimen1
      \rlap{\hbox to \dimen0{\hfil/\hfil}} #1 \else
      \rlap{\hbox to \dimen1{\hfil$#1$\hfil}} / \fi}
\ifx\oldchi\undefined \let\oldchi=\chi
  \def\cchi{{\raise 1pt\hbox{$\oldchi$}}} \let\chi=\cchi \fi

\def\frac#1#2{{\textstyle{#1 \over #2}}}

\def\half{\ifinner {\scriptstyle {1 \over 2}}\else {1 \over 2} \fi}

\def\simge{\rlap{\raise 2pt \hbox{$>$}}{\lower 2pt \hbox{$\sim$}}}
\def\simle{\rlap{\raise 2pt \hbox{$<$}}{\lower 2pt \hbox{$\sim$}}}

\def\vbig#1#2{{\vbigd@men=#2\divide\vbigd@men by 2%
\hbox{$\left#1\vbox to \vbigd@men{}\right.\n@space$}}}

%
%--------------------------------------------------------------------
%
\newcount\fdefcontre \newcount\fdefcount \newcount\indcount
\newread\filefdef  \newread\fileftmp  \newwrite\filefdef
\newwrite\fileftmp     \def\strip#1*.A {#1}
\def\futuredef#1{\beginfdef
\expandafter\ifx\csname#1\endcsname\relax%
{\immediate\write\fileftmp {#1*.A}
\immediate\write16 {Panda Warning - fdef: macro "#1" on page
\the\pageno \space undefined}
\ifnum\draftnum=0 \expandafter\xdef\csname#1\endcsname{(?)}
\else \expandafter\xdef\csname#1\endcsname{(#1)} \fi
\global\advance\fdefcount by 1}\fi   \csname#1\endcsname}

\def\beginfdef{\ifnum\fdefcontre=0
\immediate\openin\filefdef \jobname.fdef
\immediate\openout\fileftmp \jobname.ftmp
\global\fdefcontre=1  \ifeof\filefdef \immediate\write16 {Panda
WARNING - fdef: file \jobname.fdef not found, run TeX again}
\else \immediate\read\filefdef to\spzzttrra
\global\advance\fdefcount by \spzzttrra
\indcount=0      \loop\ifnum\indcount<\fdefcount
\advance\indcount by 1   \immediate\read\filefdef to\spezttrra
\immediate\read\filefdef to\sppzttrra
\edef\spzzttrra{\expandafter\strip\spezttrra}
\immediate\write\fileftmp {\spzzttrra *.A}
\expandafter\xdef\csname\spzzttrra\endcsname{\sppzttrra}
\repeat \fi \immediate\closein\filefdef \fi}
\def\endfdef{\immediate\closeout\fileftmp   \ifnum\fdefcount>0
\immediate\openin\fileftmp \jobname.ftmp
\immediate\openout\filefdef \jobname.fdef
\immediate\write\filefdef {\the\fdefcount}   \indcount=0
\loop\ifnum\indcount<\fdefcount    \advance\indcount by 1
\immediate\read\fileftmp to\spezttrra
\edef\spzzttrra{\expandafter\strip\spezttrra}
\immediate\write\filefdef{\spzzttrra *.A}
\edef\spezttrra{\string{\csname\spzzttrra\endcsname\string}}
\iwritel\filefdef{\spezttrra}
\repeat  \immediate\closein\fileftmp \immediate\closeout\filefdef
\immediate\write16 {Panda Warning - fdef: Label(s) may have changed,
re-run TeX to get them right}\fi}
\def\iwritel#1#2{\newlinechar=-1
{\newlinechar=`\ \immediate\write#1{#2}}\newlinechar=-1}
\global\fdefcontre=0 \global\fdefcount=0 \global\indcount=0
%
%--------------------------------------------------------------------
%
\null
%
%--------------------------------------------------------------------
%
%                             THE    END
%
%--------------------------------------------------------------------
%
%\input panda
%\draftmode{W-algebras.....}
\loadamsmath
\loadeuler
%
%%%%%%%%%%%%%%%%%%%%%%%%%%%%%
%
% SOME SPECIAL CHARACTERS
%
%%%%%%%%%%%%%%%%%%%%%%%%%%%%%
%
\def\hb{{\bfmath h}}
\def\ie{{\it i.e.\/}}
\def\eg{{\it e.g.\/}}
\def\gg{{\>\widehat{g}\>}}
\def\s{{\bf s}}
\def\sw{{{\bf s}_w}}
\def\Heis{{\cal H}[w]}
\def\ad{{\rm ad\;}}
\def\Ker{{\rm Ker}(\ad\Lambda)}

\def\W{$\cal W$}
%%%%%%%%%%%%%%%%%
\mathchardef\bphi="731E
\mathchardef\balpha="710B
\mathchardef\bomega="7121
\def\alb{{\bfmath\balpha}}

\def\omb{{\bfmath\bomega}}
\pageno=0
\nopagenumbers{\baselineskip=12pt
\line{\hfill US-FT/23-95}
\line{\hfill SWAT/95/83}
\line{\hfill\tt hep-th/9508084}
\line{\hfill (revised version)}
\line{\hfill July 1996}\jump
\ifdoublepage \bjump\bjump\bjump\else \vfill\fi
\centerline{\capstwo The Hamiltonian Structure of Soliton}
\jump
\centerline{\capstwo Equations and Deformed W-Algebras}
\bjump\jump
\centerline{{\scaps Carlos R.~Fern\'andez-Pousa$^{(a)}$} and {\scaps J. Luis
Miramontes$^{(a,b)}$}}
\jump
\centerline{\sl $^{(a)}$Departamento de F\'\i sica de Part\'\i 
culas,}
\centerline{\sl Facultad de F\'\i sica,}
\centerline{\sl Universidad de Santiago,}
\centerline{\sl E-15706 Santiago de Compostela, Spain}
\sjump
\centerline{\sl $^{(b)}$Department of Physics,}
\centerline{\sl University of Wales, Swansea,}
\centerline{\sl Singleton Park,}
\centerline{\sl Swansea SA2 8PP, U.K.}
\jump
\centerline{{\tt pousa@gaes.usc.es}~~ and ~~{\tt miramont@fpaxp1.usc.es}} 
\bjump
\ifdoublepage
\vfill
{\noindent
\line{July 1996\hfill}}
\eject\null\vfill\fi
\centerline{\capsone ABSTRACT}\jump

\noindent
The Poisson bracket algebra corresponding to the second
Hamiltonian structure of a large class of generalized KdV and mKdV
integrable hierarchies is carefully analysed. 
These algebras are known to have conformal properties, 
and their relation to $\cal W$-algebras has been previously investigated in
some particular cases. The class of equations that is considered
includes practically all the generalizations of the Drinfel'd-Sokolov 
hierarchies constructed in the literature. In particular, it has been recently
shown that it includes matrix generalizations of the Gelfand-Dickey and the
constrained KP hierarchies. Therefore, our results provide a unified
description of the relation between the Hamiltonian structure of soliton
equations and $\cal W$-algebras, and it comprises almost all the results
formerly obtained by other authors. The main result of this paper is an
explicit general equation showing that the second Poisson bracket algebra is a
deformation of the Dirac bracket algebra corresponding to the $\cal
W$-algebras obtained through Hamiltonian reduction.
%\bjump\bjump
%{\hfill 42 pages, 0 figures and 0 tables} 

\sjump\vfill
\ifdoublepage \else
\noindent
\line{April 1996\hfill}\fi
\eject}
\yespagenumbers\pageno=1
\footline={\hss\tenrm-- \folio\ --\hss}
%\doublespaced
%
%%%%%%%%%%%%%%%%%%%---------------!!!!!!!!!---------------%%%%%%%%%%%%
%~
%\bjump\bjump\bjump
%\noindent
%{\sl Proposed running title:}\jump
%
%\centerline{\capstwo Soliton Equations and $\cal
%W$-algebras}
%\jump
%\noindent
%{\sl Contact person:}\jump
%
%\centerline{J. Luis Miramontes Antas}
%
%\centerline{Departamento de F\'\i sica de Part\'\i culas}
%
%\centerline{Universidad de Santiago}
%
%\centerline{E-15706 Santiago de Compostela, Spain}
%
%\jump
%\centerline{Phone number: 3481-563100 ext. 14057}
%
%%\centerline{e-mail: {\tt miramont@fpaxp1.usc.es}}
%
%\newpage
%
\chapter{Introduction}

This paper completes the results of Ref.~[\Ref{WNOS}], where we investigated the
connection between the \W-algebras obtained through Hamiltonian reduction
of affine algebras~[\Ref{HAM},\Ref{BALOG},\Ref{BOU},\Ref{FREP},\Ref{FRENCH}],
and the Hamiltonian structure of the generalized Drinfel'd-Sokolov integrable
hierarchies constructed in~[\Ref{GEN1},\Ref{GEN2}].

The different hierarchies
of~[\Ref{GEN1},\Ref{GEN2}] are characterised by the data
$\{g,[w],\sw,\s,\Lambda\}$ where $g$ is a finite Lie algebra and $[w]$
indicates a conjugacy class of the Weyl group of $g$ that specifies a
Heisenberg subalgebra $\Heis$ of the affine algebra $\gg$ of
$g$. The two vectors $\sw$ and $\s$, whose components are ${\rm rank}(g)+1$
non-negative integers, define two gradations of $\gg$ such that the gradation
$\sw$ is also a gradation of $\Heis$ and $\s \preceq\sw$
(see~[\Ref{GEN1},\Ref{GEN2}]). Finally, $\Lambda$  is a constant element of
$\Heis$ with positive $\sw$-grade.  The original Drinfel'd-Sokolov
hierarchies~[\Ref{DS}] are recovered with the principal Heisenberg subalgebra,
and a remarkable property of the class of integrable hierarchies
of~[\Ref{GEN1},\Ref{GEN2}] is that it includes practically all the
generalizations of the Drinfel'd-Sokolov construction so far proposed in the
literature by several authors. 

Firstly, it includes all the hierarchies
considered in~[\Ref{WIL},\Ref{MAC}], which are recovered when the
$\sw$-grade of $\Lambda$ equals~$1$. Secondly, when the affine algebra~$\gg$ is
simply laced, it has been proved in~[\Ref{TAU}] that the integrable
hierarchies of~[\Ref{GEN1},\Ref{GEN2}] coincide with those constructed by Kac
and Wakimoto in~[\Ref{KW}] following the tau-function approach. Thirdly, the
interpretation of the integrable hierarchies of~[\Ref{GEN1},\Ref{GEN2}] by
means of pseudo-differential operators has been recently worked out
in~[\Ref{FHM},\Ref{ARA}]. To be specific, the integrable hierarchies
investigated there correspond to the conjugacy classes of the Weyl group of 
$g=sl(n,{\Bbb C})$ of the form 
$$
[w] =[\> \underbrace{r\> ,\> \ldots\> ,\> r}_{p{\;\rm times}}\> ,\>
\underbrace{1\> ,\> \ldots\> ,\> 1}_{v{\;\rm times}}\>]\>, \quad n\> =\> p\> r 
\> + \> v\>,
$$
and to the elements $\Lambda\in \Heis$ of minimal positive $\sw$-grade. Then,
when $v=0$, it is shown that the corresponding integrable
hierarchies provide $p\times p$ matrix generalizations of the Gelfand-Dickey
$r$-KdV hierarchy~[\Ref{GD}] associated to the Lax operator
$$
L_{\rm GD}\> =\> \partial^r \> +\> u_2\> \partial^{r-2}\> +\> \cdots \> +
u_{r-1}\>
\partial\> +\> u_r\>.
$$
Moreover, when $v> 0$, they lead to $p\times p$ matrix generalizations of the
constrained KP (cKP) hierarchy~[\Ref{CKP}], which is recovered for $p=v=1$ and
corresponds to the pseudo-differential Lax operator
$$
L_{\rm cKP}\> =\> \partial^r \> +\> u_2\> \partial^{r-2}\> +\> \cdots \> + 
u_{r-1}\> \partial\> +\> u_r\> +\> \phi \> \partial^{-1}\varphi\>.
$$
Finally, the link between the approach used in~[\Ref{GEN1},\Ref{GEN2}] and
the Adler-Kostant-Symes (AKS) construction has been explained
in~[\Ref{FHM},\Ref{FPLUS},\Ref{NIGEL}], where it is shown that the former
corresponds to the ``nice reductions'' of the AKS system that exhibit local
monodromy invariants. 

All this ensures that the class of integrable equations
of~[\Ref{GEN1},\Ref{GEN2}] is large enough to provide a meaningful unified
pattern of the relation between \W-algebras and the Hamiltonian structure of
soliton equations, and the purpose of this paper is to complete the results
of~[\Ref{WNOS}] about the precise form of this relation. 

In~[\Ref{WNOS}], the relation between the second Hamiltonian structure of the
integrable hierarchies of~[\Ref{GEN1},\Ref{GEN2}] and \W-algebras was
investigated in the particular case when certain non-degeneracy condition is
satisfied. Then, it was shown that the second Poisson bracket algebra
contains a \W-algebra as a subalgebra, but that only a small subset of the
\W-algebras obtained through Hamiltonian reduction are recovered from the class
of integrable hierarchies constrained by the non-degeneracy condition. These
\W-algebras are characterized by two conditions~[\Ref{WNOS}]. First, the $J_+$
that defines the associated embedding of  $sl(2,{\Bbb C})$ has to be
related in a precise way to some $\sw$-graded element $\Lambda$ of a
Heisenberg subalgebra of
$\gg$, and, second, $J_+$ has to satisfy the non-degeneracy condition. For the
different hierarchies constructed from the Lie algebra $g = sl(n,{\Bbb
C})$, these results are summarized in the Theorem~3 of~[\Ref{WNOS}], which 
shows that the non-degeneracy condition constrains the $\sw$-grade of
$\Lambda$ to be~1 or~2. It is remarkable that, using the notation of this
theorem, the cases involving $\Lambda = \Lambda^{(1)}$ coincide with the
generalizations of the constrained KP hierarchy investigated
in~[\Ref{FHM},\Ref{ARA}]. Therefore, in particular, the results of~[\Ref{WNOS}]
allow a definite identification of the \W-algebras that correspond with the
Poisson bracket algebras giving their Hamiltonian structure.

However, the non-degeneracy condition is not essential to the approach
of~[\Ref{GEN1},\Ref{GEN2}]. Indeed, it is possible to construct an
integrable hierarchy of equations from any $\sw$-graded element
$\Lambda$ of a Heisenberg subalgebra of $\gg$ that satisfies a weaker version
of that condition, which has to be required to ensure that the relevant Poisson
bracket algebra is polynomial. In this paper, we will study the Hamiltonian
structure of the integrable hierarchies obtained in the most general case when
only the weaker version of the non-degeneracy condition is satisfied.
Nevertheless, the results of~[\Ref{WNOS}] suggest that the resulting Poisson
bracket algebras can be related either to a \W-algebra or just to an affine Kac
Moody algebra (see Theorem~2). Since we are only interested in the former
case, we will still restrict our study to those hierarchies where $\Lambda$
has a non-vanishing component with zero $\s$-grade and the maximal $\s$-grade
of $\Lambda$ and of the potential $q$ equals~1 (see eq.~(5.3)). Apart from that,
the case considered in this paper will be completely general. Then, the relation
with \W-algebras will be established through the comparison of the
(second) Poisson bracket of the hierarchy with the Dirac brackets defining the
\W-algebras. 

Since our study relies on the description of the Poisson
bracket of the hierarchies as the outcome of the reduction of a Poisson
algebra, it overlaps with some results of~[\Ref{FHM},\Ref{FPLUS}]. However,
our method is closer to the spirit of the original references,
namely~[\Ref{GEN1},\Ref{GEN2},\Ref{DS},\Ref{WIL},\Ref{MAC}], and, at the end
of the day, it provides an explicit expression for the (second) Poisson
bracket algebra that relates it with certain deformations of the \W-algebras
whose properties are still to be studied.
In any case, the specialisation of our results to the cases studied
in~[\Ref{WNOS}] comfirms and clarifies the results obtained in that reference
when the non-degeneracy condition is satisfied. In particular, it shows that
the
\W-subalgebra is actually decoupled from the rest of the (second) Poisson
bracket algebra, and it allows one to investigate the existence of an energy
momentum tensor.

The paper is organized as follows. In Section~2, we briefly summarize the
required features of the integrable hierarchies of~[\Ref{GEN1},\Ref{GEN2}] and
of their second Hamiltonian structure. In particular, we point out that it is
invariant with respect to a family of conformal transformations, an important
property that is not mentioned in the original papers but which is implicit
in~[\Ref{WNOS}]. Section~3 contains an elementary review of the reduction
of Poisson manifolds, just to fix our notation. 

The next three sections constitute the body of the paper. In
Section~4, we show that the form of the second Poisson bracket follows
from the reduction of a Poisson manifold by first-class
constraints. Nevertheless, in general, the phase-space of the
integrable hierarchies is only a subset of the resulting reduced
Poisson manifold, which means that a (non-Hamiltonian in general)
additional reduction is required to obtain the second Poisson bracket
algebra. This additional reduction can be specified by selecting a
gauge slice, and a convenient choice is proposed in Section~5. This choice
generalizes the gauge slice used in~[\Ref{WNOS}] and it allows one to relate
the set of generators of the second Poisson bracket algebra with the generators
of one of the \W-algebras obtained through Hamiltonian reduction. Then, in
Section~6, we will obtain a formula, eq.~(6.18), which shows that
the second Poisson bracket algebra is given by a modification of the Dirac
bracket defining the \W-algebra; this is the main result of the paper. Whilst
Section~4 is general, in Sections~5 and~6 we only consider the restricted set
of integrable hierarchies whose Hamiltonian structures are expected to be
related to \W-algebras (see Theorem~2 of~[\Ref{WNOS}]).

In Section~7 we consider some examples that illustrate the use of eq.~(6.17),
and, in particular, our previous results in~[\Ref{WNOS}] are corroborated and
clarified. Finally, we present our conclusions in Section~8.

\chapter{Hamiltonian structure of generalized integrable
hierarchies.}
 
In~[\Ref{GEN1},\Ref{GEN2}], generalized integrable hierarchies of partial
differential equations were associated with the loop algebra $\gg =
g\otimes {\Bbb C}[z,z^{-1}]$ of a finite simple Lie algebra
$g$. Their construction requires the use of the Heisenberg
subalgebras of $\gg$, which are classified by the conjugacy classes 
of the Weyl group of $g$~[\Ref{KP}] (see also the appendix
of~[\Ref{WNOS}]).  Moreover, if $[w]$ is a conjugacy 
class and $\Heis$ is the corresponding Heisenberg subalgebra, there
exists a (non-unique) $\Bbb Z$-gradation of $\gg$, denoted by $\sw$, such that
$\Heis$ is graded by $\sw$. 

Let us remind that, up to conjugation, the
different $\Bbb Z$-gradations of $\gg$ can be defined by a set of non-negative
integers $\s=(s_0,s_1,\ldots, s_{{\rm rank}(g)})$ via the 
derivation~[\Ref{KACB}, chapter~8]
$$
d_\s\> =\> N_\s\> z\> {d\over dz} \> +\> h_\s\>,
\nfr{Derivation}
where
$$
h_\s\> =\> \sum_{j=1}^{{\rm rank}(g)} \left(2\over
\alb_{j}^2\right)\> s_j\> \omb_j\cdot\hb\>,
\quad {\rm and}\quad N_\s \>=\> \sum_{j=0}^{{\rm rank}(g)} k_j\> 
s_j\>;
\efr
in these equations, $k_0 =1$ and the $k_j$'s are the labels of the Dynkin diagram
of $g$, $\hb$ denotes a generic
element of its Cartan subalgebra, the $\alb_j$'s are the simple
roots, and the $\omb_j$'s are the fundamental weights. Then, $[h_\s,
e_{j}^\pm]= \pm s_j e_{j}^\pm$, where $e_{j}^\pm$, $j=1,\ldots,{\rm
rank}(g)$, are the raising ($+$) and lowering ($-$) operators
associated to the simple roots of $g$. Under the $\Bbb Z$-gradation
$\s$, $\gg$ decomposes as
$$
\gg=\bigoplus_{j\in{\Bbb Z}} \gg_j(\s)\>,
\qquad [\gg_j(\s), \gg_k(\s)]\subset \gg_{j+k}(\s)\>,
$$
with $\gg_j(\s) \>=\> \{v\in\gg\bigm| [d_\s\>,\> v]\>=
\>j\>v\}$. Within the set of gradations that can be
conjugated to the previous form in terms of the same basis for the
simple roots, one can introduce the following partial
ordering: $\s\preceq \s'$ if $s_{j}'\not=0$ whenever 
$s_j\not=0$, which translates into a set of inclusion
relations among the corresponding graded subspaces~[\Ref{GEN1}].

The integrable hierarchies of~[\Ref{GEN1},\Ref{GEN2}] are sets 
of zero-curvature
equations for a Lax operator constructed from the data
$\{[w],\sw,\s,\Lambda\}$, where $\s$ is a $\Bbb Z$-gradation of $\gg$
such that $\s\preceq \sw$, and
$\Lambda$ is a constant element of $\Heis$ with well defined positive
$\sw$-grade $i$, $\Lambda\in \Heis\cap \gg_i(\sw)$\note{In this paper, we
will not distinguish between regular and non-regular
$\Lambda$'s, \ie, using the terminology of~[\Ref{GEN1}], between type~I
and type~II hierarchies.}. The Lax operator takes the form
$$
L \> =\> \partial_x\> + \>\Lambda \>+ \> q(x)\>,
\nfr{LaxOp}
where the potential $q(x)$ is an element of $C^\infty({\bf
S}^1,Q)$, \ie, a periodic function\note{The choice of the domain where
the potential is defined is not crucial. The only constraint is that
its boundary conditions ensure that the integrals involved in the
definition of the Poisson brackets exist.} of $x\in {\bf S}^1$ taking
values on the subspace of $\gg$
$$
Q\> =\> \gg_{\geq0}(\s) \cap \gg_{<i}(\sw)\>;
\efr
in this last equation, we have introduced the notation $\gg_{>n}(\s)
= \bigoplus_{j>n} \gg_j(\s)$ and so on.

The function $q(x)$ plays the role of the phase-space coordinate in
this system. However, there exist symmetries corresponding to the gauge
transformations
$$
q(x) \rightarrow \widetilde{q}(x)\> =\> \exp \left(\ad S(x)\right)\>
\left(\partial_x\> + \>\Lambda \>+ \> q(x)\right)\> -\> \partial_x
\>-\> \Lambda
\nfr{GaugeTrans}
generated by any $S(x)\in C^\infty({\bf S}^1,P)$, where $P$ is the
nilpotent subalgebra
$$
P\>= \> \gg_0(\s) \cap \gg_{<0}(\sw)\>;
\nfr{GaugeGen}
the infinitesimal form of~\GaugeTrans\ is
$$
\widetilde{q}(x) \>- \> q(x) \>= \> \bigl[S(x)\>, \> \partial_x \> +
\> \Lambda\> +\> q(x) \bigr]\>, \qquad S(x) \ll1\>.
\nfr{InfGauge} 
From now on, the
group formed by these transformations will be  called $G$, and,
consequently, the actual phas- space of the system is 
$C^\infty({\bf S}^1,Q)/G$,
\ie, the set of gauge equivalence classes of Lax operators of the form~\LaxOp.

The construction of~[\Ref{GEN1},\Ref{GEN2}] is restricted to the case when the
condition\note{This condition is automatically satisfied
if $\Lambda$ is a regular element of $\Heis$~[\Ref{WNOS},\Ref{GEN1}].}
$$
\Ker \cap P \>= \> \{0\}
\nfr{Degen}
is satisfied. Actually, \Degen\ 
ensures the existence of a basis for the gauge invariant
functionals that is polynomial in the components of $q(x)$ and their 
$x$-derivatives. That basis can be easily constructed by 
following the Drinfel'd-Sokolov
procedure, which can be summarized as 
follows~[\Ref{FREP},\Ref{GEN1},\Ref{DS},\Ref{TJIN},\Ref{DSP}].
First, one has to choose some complementary space $Q^{\rm can}$ of
$[\Lambda,P]$ in $Q$, \ie, 
$$
Q\>= \> [\Lambda\>, \> P]\> + \> Q^{\rm can}\>,
\nfr{DSGF}
and, second, one simply performs a non-singular gauge
transformation to take $q(x)$ to  $q^{\rm
can}(x) \in C^\infty\bigl(S^{1}, Q^{\rm can}\bigr)$. Then, the desired
basis is provided by the components of
$q^{\rm can}(x)$ and their derivatives, understood as functionals of
$q(x)$, which implies that $C^\infty({\bf
S}^1,Q)/G$ can be identified with $C^\infty({\bf S}^1, Q^{\rm can})$.

It is convenient to distinguish the set ${\rm Pol\/}(Q)$ of 
local functionals of $q(x)$ of the form
$f[q(x)]= f\left(x, q(x), q'(x) , \ldots , q^{(n)}(x)
,\ldots\right)$ for any (differential) polynomial $f$,
and the set ${\rm Fun\/}(Q)$ of functionals of the form
$$
\varphi[q] \>= \> \int_{{\bf S}^1}\> dx\> f[q(x)]\>.
\efr
The condition~\Degen\ ensures that the phase-space of the 
hierarchy corresponds just to the subset of local gauge invariant
functionals ${\rm Pol\/}_0(Q)$ or ${\rm Fun\/}_0(Q)$, \eg,
$$
{\rm Fun\/}_0(Q)\>= \> \bigl\{\>\varphi\>\in {\rm Fun\/}(Q)\>
\bigm|\>
\varphi[\widetilde{q}]\>= \> \varphi[q]\>\bigr\} \>,
\efr
whose elements are constant on each gauge equivalence class.

One of the main properties of these  integrable hierarchies is
that they are Hamiltonian with respect to the Poisson bracket
$$
\{\varphi\>, \>\psi\}_2[q]\>= \> \biggl((d_q\varphi)_0\> ,\> 
[(d_q\psi)_0, L]\biggr)\> -\> \biggl((d_q\varphi)_{<0}\> ,\>
[(d_q\psi)_{<0}, L]\biggr)\>,
\nfr{Second}
where $\varphi, \psi \in {\rm Fun\/}_0(Q)$, $(A)_0$ and $(A)_{<0}$ 
are the components of $A\in \gg$ whose
$\s$-grade is $0$ and $<0$, respectively, and $d_q\varphi$ is the
functional derivative of $\varphi$. Even more, for the homogeneous
gradation $\s=(1,0,\ldots,0)$, the hierarchy is of the KdV type and 
it admits another coordinated Poisson bracket~[\Ref{GEN2}]
$$
\{\varphi\>, \>\psi\}_1[q]\>= \> -\biggl(d_q\varphi\> ,\> z^{-1}\>
[d_q\psi, L]\biggr)\>;
\nfr{First}
\ie, the hierarchy has a bi-Hamiltonian structure.

In~\Second\ and \First, we have used the natural invariant
non-degenerate 
bilinear form on $C^\infty({\bf S}^1,\gg)$
$$
(A,B)\> =\> \int_{{\bf S}^1} dx \langle A(x)\>, \>
B(x)\rangle_{\gg}\>,
\nfr{Bilinear}
where $\langle\cdot,\cdot\rangle_{\gg}$ is defined in terms of the
Cartan-Killing form $\langle\cdot,\cdot\rangle$ of $g$ as $\langle
a\otimes z^n, b\otimes z^m\rangle_{\gg}\> =\>
\langle a,b\rangle\> \delta_{n+m,0}$. Then, $d_q\varphi$ is
specified by the equation\note{$d_q\varphi$ is
a function of $x\in {\bf S}^1$ taking values in the
subalgebra $\gg_{\leq0}(\s)$, and it is uniquely defined by  this
equation only up to terms in $\gg_{\leq-i}(\sw)$~[\Ref{GEN2}].} 
$$
{d\over d\epsilon}\> \varphi[q\>+\> \epsilon r]\biggm|_{\epsilon=0}\>
=\> \left(d_q\varphi\>,\> r\right)\quad {\rm for\;\; all}\quad
r(x)\in C^\infty({\bf S}^1, Q)\>.
\nfr{FunDer}

In this paper, we will only be interested in the properties
of the, so called, ``second'' Poisson bracket $\{\cdot,\cdot\}_2$.
First, let us mention that both Poisson brackets are preserved by the
action of $G$, the group of gauge transformations on $C^\infty({\bf
S}^1, Q)$. Let us  consider the transformation~\GaugeTrans, and the
corresponding pullback
$\widetilde{\varphi}[q]=\varphi[\widetilde{q}]$ of a generic
functional $\varphi$. Then, it is straightforward to prove that
$\widetilde{\{\varphi , \psi\}}_{1,2} = \{\widetilde{\varphi} ,
\widetilde{\psi}\}_{1,2}$, which, in particular, ensures that the
two Poisson brackets are well defined on the set ${\rm Fun\/}_0(Q)$
of gauge invariant functionals~[\Ref{GEN2}]. The precise definition of the
second Poisson bracket within the framework of Hamiltonian reduction is
discussed in Section~4 where, in particular, it is shown that the restriction
to ${\rm Fun\/}_0(Q)$ is actually essential to ensure that this bracket is well
defined.   

The second and distinctive property of the second Poisson bracket is
that it is invariant under the conformal transformation $x \mapsto y(x)$
together with 
$$
\eqalign{
q(x)\>& =\> \sum_{j<i} q^j(x)\> e_j \cr
& \longmapsto
\breve q (y)\>= \> \sum_{j<i} (y'(x))^{j/i-1}\>
q^j(x)\> e_j\> -\> {1\over i}\>{y''(x)\over {y'}^2(x)}\> h_{\sw}\>,\cr}
\nfr{ConfTrans}
where $\{e_j\bigm| e_j\in Q\cap \gg_j(\sw)\}$ is a $\sw$-graded basis
of $Q$~[\Ref{GEN2}]\note{The equations of the integrable
hierarchy are not invariant under this conformal transformation, but
only under the scale transformations corresponding to the particular
case $y(x)=\lambda x$.}. This  conformal
transformation induces a corresponding transformation on ${\rm
Fun\/}_0(Q)$, the space of gauge invariant functionals on $Q$, which
suggests that the second Poisson bracket algebra could be a classical
extended (chiral) conformal algebra. Notice that this possibility at least 
requires the existence of an energy-momentum tensor $T(x) \in {\rm
Pol\/}_0(Q)$, such that the
infinitesimal transformation of $\varphi\in {\rm Fun\/}_0(Q)$
corresponding to $y(x) = x + \epsilon(x)$ is given by $\delta_\epsilon
\varphi = \{T_\epsilon , \varphi\}_2$ with $T_\epsilon = \int_{S^1}
dx\> \epsilon(x)\> T(x)$~[\Ref{WNOS}].

Another important property is that the two Poisson brackets admit
non-trivial centres. In fact, by construction, the $\s$-grade of the
components of $q(x)$ is bounded, and there exists a non-negative
integer $n$ such that
$$
\Lambda \>+ \> q(x)\> \in \> \bigoplus_{j=0}^n \gg_j(\s)\>.
\nfr{Bound}
Then, it follows from~\Second, and \First\ that the components of
$q(x)$ whose $\s$-grade equals $n$ are centres of the two Poisson
bracket algebras. However, in general they are not gauge invariant 
centres The existence of gauge invariant
centres has been established in~[\Ref{WNOS},\Ref{GEN2}].
For $\{\cdot,\cdot\}_2$, there may be centres only if the $\sw$-grade of 
$\Lambda$ is $i>1$ and, then, they are in one-to-one
correspondence with the elements of~[\Ref{WNOS}] 
$$
{\cal Z}\> =\> \biggl[ {\rm Ker\/}(\ad \Lambda)\cap
\gg_{i-1}(\sw)\biggr]\> \cup \> \biggl[ {\rm Cent\/}\bigl({\rm
Ker\/}(\ad \Lambda)\bigr) \cap \Bigl[ \bigoplus_{j=1}^{i-1} \gg_j(\sw)
\Bigr]\biggr]\>.
\nfr{CentreZ}
When $\Lambda$ is regular, notice that $\cal Z$
consists just of those elements of $\Heis$ whose $\sw$-grade is
$>0$ and $<i$. 

In the next sections, we will investigate the relation between
the second Poisson bracket algebra and \W-algebras
restricting ourselves to the case when $n=1$ in~\Bound. According to
the Theorem~2 of~[\Ref{WNOS}], this
restriction is required to ensure that the second Poisson bracket algebra can
be related to a \W-algebra and not just to a Kac-Moody algebra; apart from
this constraint, our results will be completely general. The non-vanishing
component of $\Lambda$ with zero $\s$-grade, $(\Lambda)_0$, is a nilpotent
element of 
$\gg_0(\s)$, and there exists $J_-\in \gg_{-i}(\sw)\cap
\gg_0(\s)$ such that $J_+ = (\Lambda)_0$, $J_-$, and $J_0=[J_+,J_-]$
close an $A_1=sl(2,{\Bbb C})$ subalgebra of
$\gg_0(\s)$~[\Ref{WNOS}]:
$$
[J_0\> ,\> J_\pm]\> =\> \pm\> J_\pm\>, \quad [J_+\> ,\> J_-]\> =\>
J_0\>;
\nfr{Embed}   
$J_0$ and $h_\s$ live in the same Cartan subalgebra of $g\simeq
g\otimes 1$ for any $\s$. Actually, this $sl(2,{\Bbb C})$ subalgebra 
specifies the \W-algebra that will be related to the second Poisson
bracket. 

Within the above mentioned restriction, the second Poisson bracket becomes 
$$
\eqalign{
\{\varphi\>, \>\psi\}_2[q]\>& = \> \biggl((d_q\varphi)_0\> ,\>
[(d_q\psi)_0\,,\partial_x + J_+ +  (q(x))_0]\biggr)\cr
& =\> \biggl([(d_q\varphi)_0\,, (d_q\psi)_0]\>, \>J_+ + 
(q(x))_0\biggr)\> + \> \biggl(\partial_x(d_q\varphi)_0\,, 
(d_q\psi)_0 \biggr)\>,\cr}
\nfr{SecondN}
which, now, is invariant under a more general class of conformal
transformations than~\ConfTrans. Let us consider 
a generic $\Bbb Z$-gradation of $\gg$, $\s^\ast$, such that 
$[J_+, h_{\s^\ast}-
i J_0]=0$, and a $\s^\ast$-graded basis $\{e_{j}^\ast\bigm|
e_{j}^\ast\in \gg_j(\s^\ast)\}$ of $Q$. Then, the second Poisson
bracket is invariant under the conformal transformation $x\mapsto
y(x)$ together with
$$
\eqalign{
(q(x))_0\>& =\> \sum_{j} {q^{\ast}}^j(x)\> e_{j}^\ast \cr
& \longmapsto
(\breve q (y))_0\>= \> \sum_{j} (y'(x))^{j/i-1}\>
{q^{\ast}}^j(x)\> e_{j}^\ast\> -\> {1\over i}\>{y''(x)\over
{y'}^2(x)}\> h_{\s^\ast}\>. \cr}
\nfr{ConfTransA}
This can be proved by following Sec.~4 of~[\Ref{GEN2}] and using that
$$
\partial_x\> +\> J_+\> +\> (q(x))_0 \>= \> y'(x) \> U[y] \Bigl(
\partial_y\> +\> J_+\> +\> (\breve q (y))_0\Bigr)\> U^{-1}[y]\>,
\efr
with $U[y] = (y'(x))^{-\>{1\over i}\> h_{\s^\ast}}$. Notice that, now,
the components of $q_1(x)$ are just centres of the second Poisson
bracket, and this is the reason why their conformal transformation is
not specified by~\ConfTransA. Actually, these centres pose the main problem in
proving the existence of an energy-momentum tensor
for~\ConfTransA~[\Ref{WNOS}].  

Therefore, the second Poisson bracket algebra has conformal properties
with respect to each of those conformal transformations and the crucial
question is whether it is a genuine $\cal W$-algebra with respect to any of
them. This means that it has to be generated by an energy-momentum
tensor and primary fields for some particular choice of $\s^\ast$,
which, according to the already known
results~[\Ref{WNOS},\Ref{FREP},\Ref{DSP}], should be the gradation induced by
$h_{\s^\ast} = iJ_0$.

\chapter{Elementary aspects of the reduction of Poisson manifolds.}   

As explained in the previous section, the condition~\Degen\ ensures
the possibility of choosing a gauge slice $Q^{{\rm
can}}\subset Q$ such that the phase-space of the integrable hierarchies
of~[\Ref{GEN1},\Ref{GEN2}] is identified with
$C^\infty({\bf S}^1,Q^{{\rm can}})$. The characteristic
properties of $Q^{{\rm can}}$ are the following: $Q=[\Lambda,P] + Q^{\rm
can}$ and, moreover, for any potential $q(x)\in C^\infty({\bf S}^1,Q)$ there
exists a unique $q(x)$-dependent gauge transformation that takes $q(x)$ to a
unique element $ q^{{\rm can}}[q(x)] \in C^\infty({\bf S}^1,Q^{{\rm
can}})$,  
$$
q(x) \>\mapsto \> q^{{\rm can}}[q(x)]\>= \>
\exp\left(\ad S^{{\rm can}}[q(x)]\right) \Bigl(\partial_x +
\Lambda + q(x)\Bigr) \>-\> \partial_x \>-\>\Lambda \>;
\nfr{Orbit}
obviously, $S^{{\rm can}}[q(x)]=0$ and $q^{{\rm can}}[q(x)] =
q(x)$ for $q(x)\in C^\infty({\bf S}^1,Q^{{\rm can}})$.
Then, if~\Degen\ is satisfied, the use of the Drinfel'd-Sokolov
procedure allows one to choose $Q^{{\rm can}}$ such that 
$S^{{\rm can}}[q(x)]$ and $q^{{\rm can}}[q(x)]$ are local
functionals of $q(x)$, \ie, elements of ${\rm Pol\/}(Q)$. Since $q^{{\rm
can}}[q(x)]$ is gauge invariant, its components and their
derivatives provide a basis for ${\rm Pol\/}_0 (Q) $ and, correspondingly, for
${\rm Fun\/}_0(Q)$.     

All this shows that there is a one-to-one map between the set 
${\rm Fun\/}_0(Q)$ of gauge invariant
functionals of $q(x)$ and the set of
functionals ${\rm Fun\/}(Q^{{\rm can}})$, and, therefore, that the second
Poisson bracket~\Second\ induces a new bracket
$\{\cdot,\cdot\}^\ast$ on ${\rm Fun\/}(Q^{{\rm can}})$ such that
the Poisson algebras $\bigl( {\rm Fun\/}(Q^{{\rm can}})\>, \>
\{\cdot,\cdot\}^\ast\bigr)$ and $\bigl( {\rm Fun\/}_0 (Q)\>, \>
\{\cdot,\cdot\}_2\bigr)$ are isomorphic. This way, all the gauge
invariant information of the original phase-space, and, in particular,
its conformal properties, is transferred to
the new Poisson algebra $\bigl( {\rm Fun\/}(Q^{{\rm can}})\>, \>
\{\cdot,\cdot\}^\ast\bigr)$.

In the next sections, it will be useful to have a formal general
definition of the induced Poisson bracket. Let $M$ be a manifold and
let $G$ be a Lie group such that each element $g\in G$ defines a
smooth action $\Phi_g : M\rightarrow M$ on the manifold $M$;
for the sake of clarity, we
will assume that $M$ is finite dimensional, even though the resulting 
expressions are valid in the general case too. We will be interested
in the case when $\bigl({\cal O},\{\cdot, \cdot\}\bigr)$ is a Poisson
algebra, where $\cal O$ is the set of $G$-invariant $C^\infty$
functions on $M$, and $\{\cdot,\cdot\}$ is a Poisson bracket in $\cal
O$. Then, let us consider the canonical projection 
$\pi:M\rightarrow M/G$ on the set of equivalence classes $M/G$. 
Since the elements
of $\cal O$ are constant on each equivalence class, its pullback map
$\pi^\ast : C^\infty(M/G) \rightarrow {\cal O}$ is actually a 
one-to-one map that assigns to any function $\hat \phi$ on $M/G$ the
$G$-invariant function $\pi^\ast(\hat \phi) = \hat \phi\circ
\pi$. This leads to the definition of the induced Poisson bracket
$$
\{ \hat\phi\>, \> \hat\psi \}^\ast\> =\> \left(\pi^\ast
\right)^{-1} \left(\Bigl\{\pi^\ast(\hat \phi)\>, \> \pi^\ast(\hat
\psi)\Bigr\} \right) 
\nfr{RPoisson}       
on $C^\infty(M/G)$, which has the property that the Poisson
algebras $({\cal O},\{\cdot,\cdot\})$ and $(C^\infty(M/G), 
\{\cdot,\cdot\}^\ast)$ are isomorphic.

Notice that, in our case, the roles
of $M$, $G$, and $\cal O$ are played by
$C^\infty({\bf S}^1, Q)$, the group of gauge transformations, and
${\rm Fun\/}_0(Q)$, respectively. 
In addition, since $C^\infty({\bf S}^1, Q)/G$ can be identified with
$C^\infty({\bf S}^1, Q^{{\rm can}})$, the canonical
projection is specified by $\pi \left( q(x)\right) = q^{({\rm
can})}[q(x)]$, and its pullback map is the isomorphism
$$
\hat \varphi[q^{\rm can}]\in {\rm
Fun\/} (Q^{\rm can})\> \longrightarrow\>  
\pi^\ast(\hat \varphi)[q] \>= \> \hat \varphi[q^{\rm can}[q]] \in {\rm
Fun\/}_0(Q)\>.
\nfr{Pullback}

In Section~5, it will be shown that $Q^{\rm can}$ can be chosen such
that it contains the set of generators of the \W-algebra associated to
the $sl(2,{\Bbb C})$ subalgebra of $\gg_0(\s)$ characterised by $J_+=
(\Lambda)_0$, \ie, all the lowest
weights corresponding to the adjoint action of the $sl(2,{\Bbb C})$
subalgebra on $\gg_0(\s)$. Then, it will be possible to define another
Poisson bracket on ${\rm Fun\/}(Q^{\rm can})$: the Dirac bracket that
especifies the \W-algebra. Since our
main objective is to relate those two, a priori different, Poisson
brackets, it will be convenient to  briefly summarize 
the main features of the
Dirac bracket. Again, for the sake of clarity, we will restrict
ourselves to the case of finite dimensional manifolds, and we will follow
the nice review included in~[\Ref{TJIN}]; more detailed discussions can be
found, for instance, in~[\Ref{PRED}].

Let us consider a Poisson manifold $\bigl({\cal M},\{\cdot,
\cdot\}\bigr)$ and a submanifold $\widetilde{M}$ that is the zero-set
of a collection of constraints $\{\phi^\mu\} \subset C^\infty({\cal
M})$, \ie,
$$
\widetilde{M} \> = \> \bigl\{ p\in {\cal M} \> \bigm|\> \phi^\mu(p) =0
\quad {\rm for\;\; all} \quad \mu\bigr\}\>.
\efr
When ${\rm det\/} \bigl(\> \overline{\{\phi^\mu, \phi^\nu\}}\> \bigr)
\not=0$, where the
bar indicates the restriction to $\widetilde{M}$, one can define the
following Poisson bracket on $C^\infty(\widetilde{M})$
$$
\{ \hat \varphi\>, \> \hat \psi \}^{\rm D}\> =\>\overline{\{
\varphi\>, \> \psi\}\> -\> \{\varphi\>, \>
\phi^\mu\}\> \Delta_{\mu\, \nu}\> \{ \phi^\nu\>, \> \psi\} }\>,
\nfr{Dirac}
which was originally considered by Dirac~[\Ref{DIRAC}] and, therefore, it is
referred to as the Dirac bracket. In~\Dirac, $\Delta_{\mu\,\nu}$ is the inverse 
matrix of $\Delta^{\mu\, \nu} = \{\phi^\mu, \phi^\nu\}$, $\hat
\varphi, \hat \psi\in C^\infty(\widetilde{M})$, and $\varphi$
and $\psi$ are two arbitrary $C^\infty$ functions on $M$ such that
$\overline{\varphi}=\hat \varphi$ and $\overline{\psi}=\hat
\psi$. Notice that the characteristic property of the right-hand-side 
of~\Dirac\ is that $\{ \cdot , \phi^\sigma\} -
\{\cdot,  \phi^\mu\} \Delta_{\mu\, \nu} \{ \phi^\nu, 
\phi^\sigma\} =0$ for any constraint $\phi^\sigma$. 

The relation between the Dirac bracket and eq.~\RPoisson\ arises 
when the reduction from $\cal M$ to
$\widetilde{M}$ is a Hamiltonian reduction by imposing first-class constraints.
Consider a submanifold $M\subset {\cal M}$ that can be given as the zero-set of
a collection of first-class constraints $\{\phi^i\} \subset C^\infty({\cal
M})$, \ie, 
$$
M\> =\> \{ p\in {\cal M} \bigm| \phi^i(p) \>= 0\>\; {\rm
for \; all}\; i\} \quad {\rm and}\quad 
\{\phi^i\>,\> \phi^j\}\biggm|_{\phi^k =0} \>=\> 0  \>.
\efr
Then, it is well known that the Poisson structure on
$\cal M$ does not induce a Poisson structure on $M$,
but on the set of equivalence classes $M/G$, where
$G$ is the group of transformations
generated by the Hamiltonian vector fields associated to the
constraints, \ie, the infinitesimal $G$-transformation generated
by $\phi^i$ is $\delta^i\> \varphi \>= \> \{\phi^i\>, \>
\varphi\}$, for any $\varphi \in C^\infty(M)$. These
transformations are usually called gauge transformations, and the
restriction to $C^\infty(M/G)$ is required to ensure that $\phi^i=0$ is
consistent  with $\{\phi^i, \cdot\}=0$ for all $\phi^i$.  
Finally, let us assume that it is possible to find
a submanifold $\widetilde{M}\subset M$, the gauge
slice, such that it has exactly one common point with every
equivalence class; then, $M/G$ can be identified with $\widetilde{M}$. 

The gauge slice $\widetilde{M}$ can be completely fixed by adding some
additional gauge fixing constraints $\{\chi^i\}$, \ie, 
$$
\widetilde{M}\> =\> \{ p\in {\cal M} \bigm| \phi^i(p) \>= \>
\chi^j(p) \> =\> 0\>\; {\rm for \; all}\; i,j\} \>.
\efr
Then, denoting by $\{\phi^\mu\}= \{\phi^i\}\cup \{\chi^j\}$ the total set
of constraints, the reduced Poisson structure on $C^\infty(\widetilde{M})$
corresponds precisely to the Dirac bracket~\Dirac\note{Properly
speaking, this is the result when $\Delta^{\mu\, \nu}$ is an
invertible matrix,
which is true if all the first-class constraints generate
transformations on $M$, and, then, the number
of first-class constrains $\phi^i$ equals the number of gauge
fixing constraints $\chi^j$. Otherwise, the first-class constraints
that do not generate any transformation on $C^\infty(M)$ can be
imposed directly, and one says that the result is a Poisson
submanifold of $\bigl({\cal M}, \{\cdot, \cdot\}\bigr)$; after doing
that, the previously described procedure can be applied to the remaining
constraints.}. One can easily check that the Dirac bracket is actually 
the Poisson bracket defined by~\RPoisson\ in this case.
Since the original constraints $\phi^i$ are first-class, $\Delta^{\mu\, \nu}$
has the block form
$$
\overline{\Delta}^{\mu\, \nu} \> =\> \overline{ \{\phi^\mu,
\phi^\nu\}}  \>=\> \bordermatrix{& \phi & \chi \cr
\phi& 0 & A \cr \chi & B & C \cr}\>,
\nfr{Block}
which means that
$$
\overline{\Delta}_{\mu\, \nu} \> =\> \bordermatrix{& \phi & \chi 
\cr \phi& -B^{-1} C A^{-1} & B^{-1} \cr \chi & A^{-1} & 0 \cr}\>.
\nfr{BlockI}
Now, for any $\hat \varphi, \hat \psi\in C^\infty(\widetilde{M})$, let
us consider the gauge invariant functions $\varphi =
\pi^\ast(\hat \varphi)$ and $\psi = \pi^\ast(\hat \psi)$, which
obviously satisfy that $\{ \varphi, \phi^i\} = \{ \psi,
\phi^i\}=0$ for all $i$. This way, taking into account~\BlockI,
the second term involved in the definition of the Dirac bracket
vanishes and
$$
\{ \hat \varphi\>, \> \hat \psi \}^{\rm D}\> =\>
\overline{\{\pi^\ast(\hat \varphi)\>, \> \pi^\ast(\hat
\psi)\}}\>= \> \overline{ \{\hat\varphi\>, \>
\hat\psi\}^\ast \circ \pi} \>= \> \{\hat\varphi\>, \>
\hat\psi\}^\ast\>,  
\nfr{Equal}
where we have used~\RPoisson, and that the restriction
of $\pi$ to ${\widetilde{M}}$ is the identity map.

\chapter{The second Poisson bracket as a reduced bracket.}

It is generally recognized that the most effective and systematic
available method to construct \W-algebras consists in the Hamiltonian
reduction of current 
algebras~[\Ref{HAM},\Ref{BALOG},\Ref{BOU},\Ref{FREP},\Ref{FRENCH}]. 
Within this method, there is a
\W-algebra associated to each embedding of $A_1=sl(2,{\Bbb C})$ into a
simple Lie algebra $g$. The components of the reduced current correspond
to the lowest weights in the decomposition of $g$ under the adjoint
action of the $sl(2,{\Bbb C})$ subalgebra, and the \W-algebra is
the Dirac bracket algebra associated to the reduction. 
In the next sections, we will establish the relation between the second
Poisson bracket algebra of the integrable hierarchies
of~[\Ref{GEN1},\Ref{GEN2}] and certain \W-algebras constructed from
$\gg_0(\s)$. Notice that, in general, $\gg_0(\s)$ is a finite
reductive Lie algebra, but the restriction of the bilinear form $\langle
\cdot, \cdot \rangle_{\gg}$ of $\gg$
provides a non-degenerate invariant bilinear form for $\gg_0(\s)$. 

We start our proof by showing that the form of 
the second Poisson bracket follows from the reduction of certain Poisson
manifold by imposing first-class constraints, which implies that the set of 
invariant functionals with respect to the group of 
transformations generated by those constraints is a Poisson manifold
with respect to the second Poisson bracket. Nevertheless, when the 
$\sw$-grade of $\Lambda$ is $i>1$, it will be apparent that the second
Poisson bracket algebra is defined only on a subset of those invariant
functionals, or, equivalently, that an additional reduction is
generally required to obtain $\bigl({\rm Fun\/}_0(Q), \{\cdot,
\cdot\}_2\bigr) \simeq \bigl({\rm Fun\/}(Q^{\rm can}), \{\cdot,
\cdot\}^\ast\bigr)$. 

Let $C^{\infty}({\bf S}^1,\gg)$ be the set of  periodic
functions $J(x)$ of $x\in {\bf S}^1$ taking values on the loop
algebra $\gg$, and let us consider the endomorphism $R_\s: \gg
\rightarrow \gg$ defined by 
$$
R_\s(A) \>= \> {1\over2} \Bigl((A)_{\geq0} - (A)_{<0}\Bigr)\>, \qquad 
A\in \gg\>.
\nfr{Rmatrix} 
$R_\s$ satisfies the modified Yang-Baxter
equation~[\Ref{GEN2},\Ref{NIGEL}], which means that it is a classical
r-matrix~[\Ref{RMAT}]. Therefore, it defines a different Lie algebra
structure on $\gg$ whose Lie bracket is
$$
\eqalign{
[A\>, \>B]_{R_\s}\> & = \> [R_\s(A)\>, \> B] \> + \> [A\> ,\> 
R_\s(B)] \cr
&=\> [(A)_{\geq0}\>, \>(B)_{\geq0}] \>- \> [(A)_{<0}\>, \>(B)_{<0}]
\>, \cr} 
\nfr{NewLie}
which satisfies the Jacobi identities as a consequence of the modified
Yang-Baxter equation.

Next, we will show that the form of the second Poisson
bracket~\Second\ follows from the reduction of the Kirillov-Poisson
bracket 
$$
\{\varphi\>, \>\psi\}_{R_\s}[J]\> =\> \Bigl([d_J\varphi\>,
\>d_J\psi]_{R_\s}\>, \> J(x)\Bigr) \> + \> \Bigl( \partial_x
(d_J\varphi)_0\>, \> (d_J\psi)_0\Bigr)\>
\nfr{KeyBrack}
by imposing first-class constraints. This Poisson bracket corresponds to the 
untwisted affinization ---in $x$--- of the infinite Lie algebra
specified by $[\cdot,\cdot]_{R_\s}$ on $\gg = g\otimes {\Bbb C}[z,z^{-1}]$,
and it is defined on the set ${\rm Fun\/}(\gg)$ of functionals of the
form $\varphi[J]= \int_{{\bf S}^1} f(x,J(x),J'(x), \ldots)$,
for any differential polynomial $f(x,J(x),J'(x), \ldots) \in
{\rm Pol\/}(\gg)$. The last term on
the right hand side of~\KeyBrack\ is a central
extension, and it is worth noticing that it can also be written
as~[\Ref{NIGEL},\Ref{RMAT}] 
$$
\Bigl(\partial_x (d_J\varphi)_0\>, \> (d_J\psi)_0 \Bigr)\>= \>
\Bigl(\partial_x R_\s(d_J\varphi)\>, \> d_J\psi\Bigr)\>+\> 
\Bigl(\partial_x d_J\varphi\>, \> R_\s(d_J\psi)\Bigr)\>.
\efr

To compare with the brief review of the previous section, notice that
$C^\infty({\bf S}^1, \gg)$ now plays the role of $\cal M$. Then, the 
space of potentials of the integrable hierarchy, $C^\infty({\bf
S}^1, Q)$, is the zero-set of the constraints
$$
\phi_\theta[J]\> =\> \left( \theta(x)\>, \> J(x) - \Lambda \right)
\> =\> \int_{{\bf S}^1} dx\> \langle \theta(x)\>, \> J(x) - \Lambda
\rangle_\gg \in {\rm Fun\/}(\gg)\>,
\nfr{Const}
for any $\theta(x) \in C^\infty({\bf S}^1, \Gamma)$, where $\Gamma$ is
the subspace
$$
\eqalign{
\Gamma\> & =\> \gg_{>0}(\s) \cup \gg_{\leq-i}(\sw) \cr
& =\> \gg_{>0}(\s) \cup \Bigl[ \gg_{<0}(\s) \cap \gg_{\leq-i}(\sw)
\Bigr] \cup \Bigl[ \gg_{0}(\s) \cap \gg_{\leq-i}(\sw) \Bigr]\>,
\cr}
\nfr{ConstForm}
\ie, $C^\infty({\bf S}^1, Q)$ is the set of functions
$q(x) = J(x) - \Lambda$ such that $\phi_\theta[\Lambda + q]=0$ for
any $\theta(x) \in C^\infty({\bf S}^1, \Gamma)$. 

Let us check that those constraints are first-class. Using that $d_J
\phi_\theta = \theta(x)$, the Poisson bracket of two constraints is 
$$
\{\phi_\theta\>, \> \phi_\gamma\}_{R_\s}[J] \>= \> \bigl( [\theta(x) ,
\gamma(x)]_{R_\s}\>, \> J(x) \bigr)\> +\> \bigl(\partial_x (\theta(x))_0\>,
\>  (\gamma(x))_0 \bigr)\>.
\efr
But, taking into account~\ConstForm, one can check that $\bigl([\theta(x) ,
\gamma(x)]_{R_\s}\bigr)_{\leq0}
\in \gg_{\leq -2i}(\sw)$, which ensures that $\bigl(
[\theta(x),\gamma(x)]_{R_\s}, \Lambda\bigr)=0$. Moreover, the last
term,  $\bigl(\partial_x (\theta(x))_0,  (\gamma(x))_0 \bigr)$ also vanishes
because both $(\theta(x))_0$ and $(\gamma(x))_0$ have $\sw$-grade
$\leq-i$. In conclusion,
$$
\{\phi_\theta\>, \> \phi_\gamma\}_{R_\s} \>= \>  \phi_{[\theta,
\gamma]_{R_\s}}\>, \quad {\rm and} \quad [\theta(x)\>,\>
\gamma(x)]_{R_\s} \in C^\infty({\bf S}^1, \Gamma)\>,
\nfr{FirstClass}  
which proves that the constraints imposed by $\phi_\theta$ with
$\theta \in C^\infty({\bf S}^1, \Gamma)$ are actually first-class.

The restriction of the Poisson-Lie bracket~\KeyBrack\ to
functions of the form $J(x) = \Lambda + q(x)$, with $q(x) \in
C^\infty({\bf S}^1, Q)$, is precisely the second Poisson bracket~\Second:
$$
\{\varphi\> ,\> \psi\}_{R_\s}[J]\biggm|_{J(x)= \Lambda + q(x)} \> = \>
\{\varphi\> ,\> \psi\}_2[q]\>, 
\efr
which, as we have explained in the previous section, is only well
defined on the set of
invariant functionals under the group of
transformations generated by the first-class constraints. 
Let $\varphi \in {\rm Fun\/}(Q)$ and $\theta(x) \in C^\infty({\bf S}^1,
\Gamma)$, then, the infinitesimal transformation of $\varphi$ generated by
the constraint $\phi_\theta$ is 
$$
\eqalign{
\delta_\theta \varphi[q] \>& = \> \{\phi_\theta\> ,\>
\varphi\}_2[q] \cr
& =\> -\Bigl( (d_q\varphi)_0\>, \> [(\theta(x) )_0 \>,\> \partial_x +
\Lambda +q(x) ]
\Bigr) \cr 
& \qquad\quad - \>
\Bigl( (d_q\varphi)_{<0}\>, \> [\Lambda +q(x) \>,\>
(\theta(x) )_{<0}] \Bigr)\>. \cr}
\nfr{TransA}
However, taking into account~\ConstForm, it follows that $[\Lambda +q(x),
(\theta(x) )_{<0}] \in \gg_{\leq0}(\sw) \subset \gg_{\leq0}(\s)$,
and, hence, the last term on the right hand side of~\TransA\ vanishes.

It is convenient to split the subspace $\Gamma$ in the
following two disjoint subsets
$$
\Gamma \> = \> \Gamma_0 \>\cup \> \Gamma_{\not=0}\>,
\nfr{SplitA}
where 
$$
\Gamma_0\> = \> \gg_{0}(\s) \cap \gg_{\leq-i}(\sw) \quad {\rm
and}\quad \Gamma_{\not=0}\>= \> \gg_{>0}(\s) \cup \Bigl[ \gg_{<0}(\s)
\cap \gg_{\leq-i}(\sw) \Bigr]\>;
\nfr{SplitB}
$\Gamma_0$ is either empty or a nilpotent subalgebra of $\gg_0(\s)$. 
Then, \TransA\ proves that ${\rm Fun\/}(Q)$ is already invariant
with respect to the transformations generated by all the constraints associated
to $\Gamma_{\not=0}$. On the contrary, when $\Gamma_0$ is not empty, 
the vector fields
$\{\phi_\theta,\cdot\}_2$ with $\theta(x) \in C^\infty({\bf S}^1,
\Gamma_0)$ generate non-trivial transformations on ${\rm Fun\/}(Q)$ that can
be understood as the infinitesimal form of
$$
q(x) \rightarrow \widetilde{q}(x)\> =\> \exp \left(\ad U(x)\right)\>
\left( \partial_x\> + \>\Lambda \>+ \> q(x) \right)\> -\> \partial_x
\>-\>, \Lambda
\nfr{ConstTrans}
for any $U(x)\in C^\infty({\bf S}^1,\Gamma_0)$. Then, the second
Poisson  bracket $\{\cdot, \cdot\}_2$ is only
well defined on those functionals which are invariant under these
transformations. 

An important observation is that $\Gamma_0$ is a subset of $P$ 
(see~\GaugeGen), but $\Gamma_0 \not= P$ whenever $i>1$. This means that,
in general, the group of transformations generated by the first-class
constraints~\Const\ is only a subgroup of the group of gauge
transformations of the integrable hierarchy, and, consequently, the
phase-space of the integrable hierarchy is actually a subset of the
set of functionals which are left invariant by those first-class
constraints. Therefore, our discussion in the previous
paragraph ensures that the second Poisson bracket algebra is
well defined\note{This result is equivalent to the lemma 3.2
of~[\Ref{GEN2}], and it is interesting to compare the proof presented there
with ours.}. Nevertheless, when $i>1$ and, consequently, $P\not=
\Gamma_0$, it also shows that the second Poisson bracket algebra
is the restriction of $\{\cdot,\cdot\}_2$ to ${\rm
Fun\/}_0(Q)$, which means that an additional reduction is required to obtain
$\{\cdot,\cdot\}^\ast$. 

The different roles of the two subsets of constraints associated
to $\Gamma_{\not=0}$ and $\Gamma_0$ suggest to describe the
reduction leading from~\KeyBrack\ to~\Second\ as a two
steps process. The first step would be the reduction
of~\KeyBrack\ with the first-class constraints associated to
$\Gamma_{\not=0}$, which leads to a reduced current of the form
$$
J(x)\> -\> \Lambda \in \gg_0(\s) \cup \bigl[ \gg_{>0}(\s)
\cap \gg_{<i}(\sw)\bigr]\>.
\nfr{RedA}
Notice that there is no restriction at all on the components of
$J(x)$ in the subalgebra $\gg_0(\s)$. Therefore, only the
component of $\Lambda$ whose $\s$-grade is positive is relevant at
this step, and~\RedA\ is equivalent to $J(x) = (\Lambda)_{>0} +
\kappa (x)$, where $\kappa (x)\in C^\infty({\bf S}^1, Q^\bullet)$ with
$$
Q^\bullet \> =\> \gg_0(\s) \cup \bigl[\gg_{>0}(\s)
\cap \gg_{<i}(\sw)\bigr]\>.
\efr
Since the first-class constraints associated to $\Gamma_{\not=0}$ do
not generate any transformation on the reduced manifold, the
bracket~\KeyBrack\ induces a well defined Poisson structure on
$C^\infty({\bf S}^1, Q^\bullet)$ and, correspondingly, on the set of
functionals ${\rm Fun\/}(Q^\bullet)$:
$$
\eqalign{
\{\varphi, \psi\}^\bullet [\kappa] \> & =\> \{\varphi, \psi\}_{R_\s} [J]
\biggm|_{J(x)= (\Lambda)_{>0} + \kappa(x)}\>= \>
\Bigl((d_{\kappa}\varphi)_0\>,
\>
\bigl[ (d_{\kappa}\psi)_0, \partial_x + (\kappa(x))_0 \bigr] \Bigr)
\cr & \qquad -\> \Bigl((d_{\kappa}\varphi)_{<0}\>, \> \bigl[
(d_{\kappa}\psi)_{<0}, (\Lambda + \kappa(x))_{>0} \bigr] \Bigr)
\>; \cr}
\nfr{BraPart}
in other words, $\bigl({\rm
Fun\/}(Q^\bullet), \{\cdot,\cdot\}^\bullet
\bigr)$ is a Poisson submanifold of $\bigl({\rm Fun\/}(\gg), 
\{\cdot,\cdot\}_{R_{\s}} \bigr)$.

Once we have imposed the constraints associated to
$\Gamma_{\not=0}$, the second step is the reduction of~\BraPart\ with
the first-class constraints induced by $\Gamma_0$. Actually, this
reduction is independent of the previous one, which suggests that
$(\Lambda)_0$ could be generally chosen independently of
$(\Lambda)_{>0}$. Instead, in order to recover the second Poisson
bracket of the integrable hierarchies of~[\Ref{GEN1},\Ref{GEN2}], it is 
constrained by the condition that $\Lambda =(\Lambda)_0 +
(\Lambda)_{>0}$ is a constant graded element of $\Heis\cap \gg_i(\sw)$.
Then, $\kappa(x) = (\Lambda)_0 + q(x)$, and the resulting reduced bracket is
the second Poisson bracket~\Second, which is only well defined on the set of
invariant functionals with respect to the transformations~\ConstTrans. Since
$\gg_{\leq0}(\sw)\subset \gg_{\leq0}(\s)$, this group of
transformations has the characteristic property that
$$
\exp \bigl(\ad U(x)\bigr)\>
\left( (\Lambda)_{>0} \right)\> = \> (\Lambda)_{>0}\>
\nfr{LambCond}
for any $U(x)\in C^\infty({\bf S}^1, \Gamma_0)$. This has already
been realized by the authors of~[\Ref{FHM},\Ref{FPLUS}], where they suggest that
new reductions of affine algebras could be obtained by enlarging this group
of transformations while keeping the condition~\LambCond.

\chapter{A convenient choice for the gauge slice $Q^{\rm can}$.}

So far, we have exhibited the relation between the second Poisson
bracket of the integrable hierarchies of~[\Ref{GEN1},\Ref{GEN2}] and the 
reduction of
the Kirillov-Poisson bracket~\KeyBrack\ by imposing first-class
constraints. The result is that
the second Poisson bracket is actually well defined for those
functionals of $q(x)$ that are invariant under the gauge transformations
generated by the elements of $C^\infty(S^1, \Gamma_0)$; 
nevertheless, the phase-space of the integrable hierarchies consists only of
the gauge invariant functionals under the transformations generated by
the elements of $C^\infty(S^1, P)$, and
$$
\Gamma_0 \>= \> \gg_{0}(\s) \cap \gg_{\leq -i}(\sw)\> 
\subseteq \> P \>= \> \gg_{0}(\s) \cap \gg_{<0}(\sw)\>.
\nfr{CompGauge}

In the particular case when the $\sw$-grade of $\Lambda$
is $i=1$, $\Gamma_0$ and $P$
coincide. Then, the second Poisson bracket
algebra can be entirely understood in terms of first-class constraints
and, according to Section~3, the new bracket induced on the
set ${\rm Fun\/}(Q^{\rm can})$ is the associated Dirac
bracket; \ie, the new bracket corresponds to a \W-algebra. 
Actually, when $i=1$ notice that $[\Lambda, P] \in \gg_0(\s)$ and $P=\Gamma_0$,
which implies that
$$
[\Lambda\>, \>P]\> = \>[(\Lambda)_0 \>,
\> P]\>,
\efr
Therefore, the condition~\Degen\ is equivalent to the non-degeneracy
condition and, hence, this case is one of those considered in~[\Ref{WNOS}]. It
is important to remark that most of the results
of~[\Ref{FHM},\Ref{ARA},\Ref{FPLUS}] are restricted to integrable hierarchies
constructed from elements $\Lambda$ whose $\sw$-grade equals 1. 

In contrast, our objective is to investigate the structure of the
new bracket $\{\cdot,\cdot\}^\ast$ in the general
situation when $i>1$. Nevertheless, we are only interested in those integrable
hierarchies whose second Poisson bracket algebra is expected to be related to a
\W-algebra and not just to a Kac-Moody algebra; consequently, and according to
the Theorem~2 of~[\Ref{WNOS}], we will only consider integrable hierarchies
where 
$$
\Lambda\>+ \> q(x) \in \gg_0(\s) \oplus \gg_1(\s)\>,
\nfr{NOne}
and $(\Lambda)_0 \not=0$, which, 
in particular, implies that $\Gamma_0$ is not empty.

Then, $(\Lambda)_0$ is a nilpotent element of
$\gg_0(\s)$, and the Jacobson-Morozov theorem~[\Ref{MORZ}] 
affirms that there exists an
element $J_-\in \gg_0(\s)\cap \gg_{-i}(\sw)$ such that
$J_+=(\Lambda)_0$, $J_-$, and $J_0 =[J_+,J_-]$ generate an
$sl(2,{\Bbb C})$ subalgebra of $\gg_0(\s)$. Moreover, $J_0$ and
$h_{\sw}$ live in the same Cartan subalgebra of $g$ 
(properly speaking, of $g\otimes 1$), and they are related in such a
way that $Y= h_{\sw} - iJ_0$ commutes with $J_\pm$~[\Ref{SORB}]; actually,
$J_0$, $J_\pm$, and $Y$ generate a $sl(2,{\Bbb
C}) \oplus u(1)$ subalgebra of $\gg_0(\s)$. Under the adjoint
action of this subalgebra, $\gg_0(\s)$ decomposes as the
direct sum of a finite number of irreducible representations
$$
\gg_0(\s) \>= \> \bigoplus_{k=1}^{n} D_{j_k} (y_k)\>,
\nfr{Repr}
where $j_k$ and $y_k$ are the $sl(2,{\Bbb C})$ spin and the
eigenvalue of $Y$ that label each representation. In addition, the
decomposition~\Repr\ ensures that $\gg_0(\s)$ has the following two
orthogonal decompositions with respect to the bilinear form
$\langle\cdot, \cdot\rangle_\gg$\note{From now on, it will be assumed that the
operators
$(\ad J_\pm)$ are restricted to
$\gg_0(\s)$, even  though we will not explicitly indicate it.}
$$
\eqalign{
\gg_0(\s) \> & =\> {\rm Ker}(\ad J_+) \oplus {\rm Im}( \ad J_-)\cr
& =\> {\rm Ker}(\ad J_-) \oplus {\rm Im}( \ad J_+)\>, \cr}
\nfr{Weights}
where ${\rm Ker}(\ad J_\pm)$ is the subset of highest $(+)$
or lowest $(-)$ weights with respect to the $sl(2,{\Bbb C})$
subalgebra. Since $Y=h_{\sw} -i J_0$ commutes with $J_\pm$, all the
subspaces ${\rm Ker\/}(J_\pm)$ and ${\rm Im\/}(J_\pm)$ are stable under
the adjoint action of $h_\sw$ and, hence, they can be decomposed under
the ${\Bbb Z}$-gradation $\sw$.

Recall that $J_0$ also induces a ${\Bbb Z}/2$-gradation of $\gg_0(s)$ where
the elements of $D_{j_k}(y_k)$ have grades $\{-j_k,\> -j_k+1,\>
\ldots, \>j_k-1,\> j_k\}$, while their corresponding $\sw$-grades are
$\{-ij_k+y_k,\> -i(j_k-1) +y_k,\> \ldots, \>i(j_k-1)+y_k,\> ij_k +y_k\}$.
However, in general, it is not possible to compare the graded subspaces of
$\gg_0(\s)$ corresponding to this gradation and to $\sw$. Let $i_1,\ldots,i_p$
be all the indices for which $s_{i_1}=\cdots =s_{i_p} =0$, \ie, the
set of vanishing components of $\s$. Then, $\gg_0(\s)$ is the
reductive finite Lie algebra 
$$
\gg_0(\s)\> =\> h_1 \oplus \cdots\oplus  h_q \oplus u(1)^{\oplus ({\rm
rank}(g) - p)}\>,
\efr
where $h_1 \oplus \cdots \oplus h_q$ is the semisimple Lie algebra whose
Dynkin diagram is the subdiagram of the extended Dynkin diagram of
$g$ consisting of the vertices $i_1,\ldots,i_p$.
Then, the problem is that the raising (lowering)
operators associated to the vertices $i_1, \ldots, i_p$ do not always
have non-positive (non-negative) grade in the gradation induced by $J_0$.
For example, this means that one cannot ensure that ${\rm
Ker\/}(\ad J_-)$ is a subset of $\gg_{\leq0}(\sw)$, even though the grade of
the elements of ${\rm Ker\/}(\ad J_-)$ is $\leq0$ in the gradation
induced by $J_0$.

Concerning the potential, eq.~\NOne\ implies that it can have non
vanishing components in
$C^\infty({\bf S}^1, Q\cap\gg_1(\s))$ when $i>1$, \ie,
$$
q(x) \> =\> (q(x))_0 \> +\> (q(x))_1\>, 
\efr
Nevertheless, even if $i>1$, there will be cases when this component is
absent. For instance, if $\s=(1,0,\ldots,0)$ is the homogeneous
gradation and $i\leq (\sw)_0$ then $(q(x))_1=0$;
actually, in these cases, $i= (\sw)_0$ is the lowest possible
positive $\sw$-grade of the elements of $\Heis$~[\Ref{WNOS}]. 
From now on, we will also use the notation
$$
\Lambda \>=\> J_+\> +\> \lambda_1\>, \;\; {\rm where}\;\;
\lambda_1\in
\gg_1(\s)\cap \gg_{i}(\sw)\>.
\efr

When~\NOne\ is fulfilled, the second Poisson bracket simplifies
to~\SecondN, which resembles the Kirillov-Poisson bracket
corresponding to the untwisted affinization ---in $x$--- of the
finite Lie algebra
$\gg_0(\s)$, where the affine $\gg_0(\s)$ current has been reduced to
$J(x)= J_+ + (q(x))_0$. The difference is that, in~\SecondN, 
$\varphi$ and $\psi$ are gauge invariant functionals not only of $(q(x))_0$, 
but also of $
(q(x))_1$ when $i>1$. Then, even though the components of $(q(x))_1$
are centres of $\{\cdot ,\cdot\}_2$, the (infinitesimal) gauge
invariance condition for $\varphi\in {\rm Fun\/}_0(Q)$ (see~\InfGauge)
$$
\eqalign{
\delta_S \varphi\> &= \>\varphi[\widetilde{q}] - \varphi[q] \>= \>
\left( d_q \varphi\>, \> [S(x), \partial_x + \Lambda + q(x)] \right)
\cr
& =\> \left( (d_q \varphi)_0\>, \> [S(x), \partial_x + J_+ + (q(x))_0]
\right)
\> \cr
&\quad\qquad +\> \left( (d_q \varphi)_{-1}\>, \> [S(x), \lambda_1 + (q(x))_1]
\right)\> = \> 0\cr}
\nfr{GaugeInvar}
relates the dependence on $(q(x))_0$ and on $(q(x))_1$.

As we have explained in Section~2, the phase-space of the integrable
hierarchies of~[\Ref{GEN1},\Ref{GEN2}] is the set ${\rm Fun\/}_0(Q)$ of gauge
invariant functionals. In this case, the gauge
transformations~\GaugeTrans\ can be expressed as
$$
\eqalign{
& (q(x))_0 \rightarrow (\widetilde{q}(x))_0\> =\> \exp \left(\ad
S(x)\right)\> \left(\partial_x\> + \>J_+ \>+ \> (q(x))_0\right)\> -\>
\partial_x \>-\> J_+\>, \cr
&(q(x))_1 \rightarrow (\widetilde{q}(x))_1\> =\> \exp \left(\ad
S(x)\right)\>
\left(\lambda_1 \>+ \> (q(x))_1\right)\> -\> \lambda_1\>, \cr}
\nfr{GTExp}
for any $S(x)\in C^\infty({\bf S}^1,P)$. Moreover, $\Lambda$ is
restricted by the condition~\Degen, which ensures the possibility of
choosing a gauge slice $Q^{{\rm can}}\subset Q$ such that the set of
equivalence classes $C^\infty({\bf S}^1,Q)/G$ and the set of gauge
invariant functionals ${\rm Fun\/}_0(Q)$ can be identified with
$C^\infty({\bf S}^1,Q^{{\rm can}})$ and ${\rm Fun\/}(Q^{{\rm can}})$,
respectively (see eq.~\Pullback). 

The condition~\Degen\ is a weak version of the 
non-degeneracy condition
$$
{\rm Ker\/}(\ad J_+) \cap P\> =\> \{0\}
\nfr{DegenPlus}
that is required in the Hamiltonian reduction
approach to $\cal W$-algebras to ensure polynomiality~[\Ref{FREP},\Ref{DSP}],  
and it plays exactly the same role here. Actually,
consider the subset
$\Gamma_0 = P\cap \gg_{\leq-i}(\sw)$ of $P$. Since $\gg_{\leq0}(\sw)
\subset
\gg_{\leq0}(\s)$, the condition~\Degen\ implies that ${\rm Ker}(\ad J_+)
\cap \Gamma_0 = \{0\}$, which means that all the elements of
$\Gamma_0$  satisfy the non-degeneracy condition~\DegenPlus. Then,
the weaker condition~\Degen\ permits that some elements of $P\cap
\gg_{>-i}(\sw)$ commute with $J_+$ only if they do not also commute with
$\lambda_1$; therefore,~\Degen\ can also be expressed as
$$
\bigl[ {\rm Ker}(\ad J_+) \cap P \bigr]\cap
\bigl[ {\rm Ker}(\ad \lambda_1) \cap P \bigr]\> = \> \{0\}\>.
\efr

Another important consequence of~\Degen\ is the following. Since the
non-degenerate invariant bilinear form of $\gg$ provides a one-to-one
map between ${\rm
Ker}(\ad J_+) $ and ${\rm Ker}(\ad J_-) $, and between $\gg_{\leq-i}$
and $\gg_{\geq i}$, the condition~\Degen\ implies that
$ {\rm Ker}(\ad J_-)\cap \gg_{\geq i} = \{0\}$,
which ensures that all the lowest weights in the decomposition~\Repr\
are included in the subspace $Q$:
$$
{\rm Ker}(\ad J_-)\> \subset\>  Q\>.
\nfr{NoLowest}

The study of the Poisson bracket induced on ${\rm Fun\/}(Q^{{\rm
can}})$ requires a convenient choice of the gauge slice, which will
be achieved through the Drinfel'd-Sokolov 
procedure~[\Ref{FREP},\Ref{GEN1},\Ref{DS},\Ref{TJIN},\Ref{DSP}]. 
In the following, we will show that the gauge slice can be chosen such that
$Q^{{\rm can}} \cap \gg_0(\s)= {\rm Ker}(\ad J_-)$, which is very convenient
since $C^\infty({\bf S}^1,{\rm Ker}(\ad J_-))$ is the phase-space of one of the
\W-algebras obtained through the Hamiltonian reduction of the affine Lie 
algebra of
$\gg_0(\s)$~[\Ref{HAM},\Ref{BALOG},\Ref{BOU},\Ref{FREP},\Ref{FRENCH}]. 

We start by decomposing the subalgebra $P$ of gauge
transformation generators as
$$
P\> =\> \overline{P}\cup P^{\star}\>,
\nfr{PPartA}
such that
$$
P^{\star}\> =\> {\rm Ker}(\ad J_+) \cap P\>, \qquad{\rm
and}\qquad {\rm Ker}(\ad J_+) \cap \overline{P} \> =\> \{0\}\>;
\nfr{PPartB}
obviously, $\overline{P}\cap P^{\star} = \{0\}$. The subset $P^\ast$ contains
those elements that do not satisfy the non-degeneracy
condition~\DegenPlus, and it is always empty for~$i=1$; moreover, the
set $\Gamma_0$ is always a subset of $\overline{P}$.
Now, let us consider the gauge transformation
generated by an element $\alpha_{(-j)}(x) \in C^\infty({\bf S}^1,P\cap
\gg_{-j}(\sw))$. According to~\PPartA, this element can be decomposed
as
$\alpha_{(-j)}(x) = \overline{\alpha}_{(-j)}(x) + \alpha_{(-j)}^\star(x)$,
where $\overline{\alpha}_{(-j)}(x) \in C^\infty({\bf S}^1,\overline{P})$
and $\alpha_{(-j)}^\star(x) \in C^\infty({\bf S}^1,P^{\star})$, and the
gauge transformation is just
$$
\eqalignno{
& (\widetilde{q}(x))_0\> =\> (q(x))_0\> - \> [J_+,
\overline{\alpha}_{(-j)}(x)] \> +\> \cdots  \>, &\nameali{FixZ}\cr
&(\widetilde{q}(x))_1\> =\> \biggl((q(x))_1\> - \> [\lambda_1,
\overline{\alpha}_{(-j)}(x)]\biggr) \> -\> [\lambda_1,
\alpha_{(-j)}^\star(x)] \>+ \> \cdots \>, &\nameali{FixO}\cr}
$$
where the dots indicate terms whose $\sw$-grade is $<i-j$. Notice
that this transformation does not change the components of
$q(x)$ whose grade is $>i-j$. Then, considering~\Weights\ and
\NoLowest, it immediately follows that
$\overline{\alpha}_{(-j)}(x)$ can be fixed uniquely with eq.~\FixZ\ by the
condition that the component of $(\widetilde{q}(x))_0$ whose
$\sw$-grade equals $i-j$ lives in ${\rm Ker}(\ad J_-)\cap
\gg_{i-j}(\sw)$. Once $\overline{\alpha}_{(-j)}(x)$ is known,
$\alpha_{(-j)}^\star(x)$ can be fixed using eq.~\FixO\ and some
appropriate choice of $Q^{\rm can}\cap \gg_1(\s)$.

If we denote by $-p$ the
lowest $\sw$-grade of the elements of $\gg_0(\s)$,
all this shows that there exist unique elements $\alpha_{(-1)}(x),
\ldots, \alpha_{(-p)}(x) \in C^\infty({\bf S}^1,P)$ such that
$$
\exp\left(\ad \alpha_{(-p)}(x)\right)\circ \cdots \circ \exp\left(\ad
\alpha_{(-1)}(x)\right) \> \equiv \> \exp \left(\ad S^{{\rm can}}(x) \right) 
\efr 
generates a gauge transformation $q(x) \rightarrow \widetilde{q}(x)
\> =\> q^{{\rm can}}(x)$, where $ q^{{\rm can}}(x)\in C^{\infty}({\bf
S}^1, Q^{\rm can})$ and (see~\NoLowest)
$$
Q^{{\rm can}} \cap \gg_0(\s)\> =\> {\rm Ker}(\ad J_-)\>.
\efr
The consequence of using the Drinfel'd-Sokolov procedure is that the
components of $S^{{\rm can}}(x)$ and $q^{{\rm can}}(x)$ are local functionals
of $q(x)$ and, in particular, $q^{{\rm can}}(x)$ is a local gauge invariant
functional, \ie,
$$
S^{{\rm can}}(x)\>=\> S^{{\rm can}}[q(x)]\in {\rm Pol\/}(Q)\>,
\qquad q^{{\rm can}}(x) \> = \> q^{{\rm can}}[q(x)]
\in {\rm Pol\/}_0(Q)\>.
\efr

A very important feature of this choice of the gauge slice is
that $Q^{\rm can}\cap \gg_0(\s)$ is completely specified by the
$sl(2,{\Bbb C})$ subalgebra $(J_0,J_\pm)$, \ie, by $(\Lambda)_0$;
hence, it is somehow independent of the
gradation $\sw$ we have started with. This is important
since, at the end of Section~2, we have already pointed out that 
there are different conformal transformations compatible with the second
Poisson bracket, and that they are associated with those gradations 
$\s^\ast$ such that
$[(\Lambda)_0, h_{\s^\ast} - i J_0]=0$; from this point of view, $\sw$
is just a particular choice of $\s^\ast$.

\chapter{The second Poisson bracket as a modified Dirac bracket.}

As explained in Section~3, the restriction of the
second Poisson bracket to the gauge invariant functionals on $Q$
specifies a new bracket $\{\cdot,\cdot\}^*$ in ${\rm Fun\/}(Q^{\rm
can})$ such that the Poisson manifolds
$({\rm Fun\/}_0(Q), \{\cdot,\cdot\}_2)$ and $({\rm Fun\/}(Q^{\rm
can}), \{\cdot,\cdot\}^*)$ are isomorphic. The new bracket is
formally defined in~\RPoisson\ where, in this case, the map is 
$$
\pi : C^\infty({\bf S}^1,Q) \rightarrow C^\infty({\bf S}^1, Q^{\rm
can})\>, \qquad \pi(q(x))\>=\> q^{{\rm can}}[q(x)]\>,
\efr
whose pullback is given by eq.~\Pullback. In the previous section, we
have shown that the gauge slice $Q^{\rm can}$ can be chosen such
that it contains the  phase-space of the 
\W-algebra specified by the embedding of $sl(2,{\Bbb C})$
into $\gg_0(\s)$ corresponding to $J_+= (\Lambda)_0$,
$$
Q^{\rm can} \> = \> {\rm Ker\/}(\ad J_-) \> \oplus \> [Q^{\rm can}
\cap \gg_1(\s)]\>.
\efr
Our next objective is to establish a precise relation between the
bracket $\{\cdot ,\cdot\}^\ast$ and this \W-algebra.

According to Section~4, the Poisson manifold $({\rm Fun\/}(Q^{\rm
can}), \{\cdot,\cdot\}^*)$ is a
reduction of $({\rm Fun\/}(Q^\bullet), \{ \cdot, \cdot\}^\bullet)$
(see eq.~\BraPart), with
$$
\{\varphi, \psi\}^\bullet [\kappa] \> =\>
\Bigl((d_{\kappa}\varphi)_0\>,
\>
\bigl[ (d_{\kappa}\psi)_0, \partial_x + (\kappa(x))_0 \bigr]
\Bigr)\>,
\nfr{BraBul}
and $Q^\bullet = \gg_0(\s) \cup [\gg_1(\s)\cap \gg_{<i}(\sw)]$.
The set of functions $C^\infty({\bf S}^1, Q^{\rm can})$
can be expressed as the zero-set of the following functionals
$$
\phi_\theta[\kappa] \> =\> \Bigl(\theta(x), \kappa(x)-
(\Lambda)_0\Bigr)\in {\rm Fun\/}(Q^\bullet)\>,
\efr
where, according to~\Weights, $\theta(x)$ is a function of $x$ 
taking values on
$$
\Gamma^{\rm can}\> =\> {\rm Im\/}(\ad J_-) \cup
\Upsilon\>, 
\efr
and $\Upsilon$ is a subset of $\gg_{-1}(\sw)\cap \gg_{\geq-i}(\sw)$ that
specifies the form of $Q^{\rm can}\cap\gg_1(\s)$. 

The constraints associated to $\Upsilon$ can be imposed directly
because they do not generate any transformation on the reduced
manifold, which is always a Poisson submanifold of $({\rm Fun\/}(Q^\bullet), 
\{\cdot, \cdot\}^\bullet)$; this is the reason why we have not indicated any
particular choice for $Q^{\rm can}\cap\gg_1(\s)$. Then, we are left only with
the constraints associated to $\Gamma^{\rm can}\cap \gg_0(\s)$, which can be
split in the following three disjoint subsets
$$
\eqalignno{
\Gamma^{\rm can}\cap \gg_0(\s)\>& =\> \Gamma_0 \cup \Gamma_1 \cup
\Gamma_2\>, \cr
\noalign{\vskip0.2cm}
\Gamma_0 \>& = \> {\rm Im\/}(\ad J_-)\cap \gg_{\leq-i}(\sw)\cr
\Gamma_1 \>&= \> {\rm Im\/}(\ad J_-)\cap \Bigl[\gg_{>-i}(\sw) \cap
\gg_{<0}(\sw) \Bigr]\cr
\Gamma_2 \>&= \> {\rm Im\/}(\ad J_-)\cap \gg_{\geq0}(\sw)\>; &
\nameali{SplitFin} \cr} 
$$
notice that the two definitions of $\Gamma_0$ in eqs.~\SplitB\ and
\SplitFin\ coincide because its elements satisfy the non-degeneracy
condition~\DegenPlus. Moreover, the non-degenerate invariant bilinear form of
$\gg$ provides a one-to-one map between $\Gamma_2$ and ${\rm
Im\/}(\ad J_+)\cap \gg_{\leq0}(\sw)=[J_+, \Gamma_0]$. Using again the
fact that the elements of $\Gamma_0$ satisfy the non-degeneracy
condition, this identification provides a one-to-one map between
$\Gamma_0$ and $\Gamma_2$, which, in particular, shows that they
have the same dimension as vector subspaces.

Within the Hamiltonian reduction construction of \W-algebras, they are
given in terms of Dirac brackets. In our case, we can construct the
Dirac bracket corresponding to the Hamiltonian reduction from $({\rm
Fun\/}(Q^\bullet),  \{ \cdot, \cdot\}^\bullet)$ to $({\rm Fun\/}(Q^{\rm
can}), \{\cdot,\cdot\}^D)$. To do that, it will be convenient to
introduce $x$-dependent constraints. Let us choose a basis $\{\theta^j\}$ for
$\Gamma^{\rm can}\cap \gg_0(\sw)$ and consider the constraints
$$
\phi^i(x) \>=\> \Bigl\langle \theta^i\>, \> (\kappa(x))_0 - 
(\Lambda)_0\Bigr\rangle_{\gg}\in {\rm Pol\/}(Q^\bullet)\>,
\quad{\rm for\;\; any}\quad x\in S^1\>.
\nfr{ConsLoc}
Then, the Dirac bracket is given by the field theoretical version of~\Dirac
$$
\eqalign{
\{\hat \varphi\>, \> \hat\psi\}^D \> &= \>
\overline{\{\varphi\>,\> \psi\}_2} \cr 
& - \> \sum_{i,j}\>
\int_{S^1} \> dx\> dy\> \overline{\{\varphi\>,
\>\phi^i(x)\}_2}\>\>
\Delta_{i,j}(x,y)\>\>
\overline{\{\phi^j(y)\> ,\> \psi\}_2}\>, \cr}
\nfr{DiracF}
where $\Delta_{i,j}(x,y)$ is the inverse of
$$
\eqalign{
\Delta^{i,j}& (x, y)\> [q^{\rm can}]\>  =\> 
\overline{\{ \phi^i(x)\>, \> \phi^j(y)\}_2 [q]} \cr
& =\> \Bigl\langle [\theta^i, \theta^j]\>,\> J_+ + (q^{\rm can}(x))_0
\Bigr\rangle_{\gg}\> \delta(x-y) \>+ \>
\Bigl\langle \theta^i\> , \>\theta^j \Bigr\rangle_{\gg} \partial_x 
\delta(x-y)\>, \cr}
\nfr{OpMat}
\ie,
$$
\sum_{k} \> \int_{S^1} dw\> \Delta_{i,k}(x,w)\> \Delta^{k,j} (w,y)\> =\> 
\delta_{\>i}^{\>j}\>  \delta(x-y)\>,
\efr
and, in general, it is a matrix differential operator.
In the last equations, the bar denotes the restriction from 
$Q^\bullet$ to the
gauge slice $Q^{\rm can}$. This restriction can be
done in two steps. First, from $Q^\bullet$ to $Q$, which means that 
$\kappa(x) -
(\Lambda)_0 =  q(x) \in C^\infty(S^1, Q)$, and, second, from
$Q$ to $Q^{\rm can}$, \ie, $q(x) \in C^\infty(S^1, Q^{({\rm
can})})$. This has been taken into account in~\DiracF\ and \OpMat,
where we have also used that 
$$
\{\cdot, \cdot\}^\bullet[\kappa]\Bigm|_{\kappa =(\Lambda)_0 + q} \> =\>
\{\cdot, \cdot\}_2[q]\>.
\efr
Moreover, $\hat \varphi,\hat\psi \in {\rm Fun\/}(Q^{\rm can})$, and $\varphi$
and $\psi$ are any two functionals in ${\rm Fun\/}(Q)$ such that
$\overline{\varphi}=\hat \varphi$ and $\overline{\psi} = \hat \psi$.  

In particular, to calculate the Dirac bracket, one can choose
$\varphi=\hat\varphi$ and $\psi = \hat \psi$. Then, since the components of
$(q(x))_1$ are centres of both $\{\cdot,\cdot\}^\bullet$ and
$\{\cdot,\cdot\}_2$, and since $Q^{\rm can}$ is a subset of $Q$, it
follows that all the components of $(q^{\rm can}(x))_1$ are
just centres of $\{\cdot,\cdot\}^D$. Therefore, the Dirac bracket~\DiracF\ 
is non-degenerate only when it is restricted to
${\rm Fun\/}(Q^{\rm can})\cap \gg_0(\s)$, and it corresponds just to the
reduction of $\bigl({\rm Fun\/}(\gg_0(\s)), \{\cdot, \cdot\}^\bullet\bigr)$,
which is a classical realization of the centrally extended affine current
algebra of $\gg_0(\s)$, to
$\bigl({\rm Fun\/}({\rm Ker\/}(\ad J_-)), \{\cdot, \cdot\}^D\bigr)$. All this
proves that the later is the \W-algebra corresponding to $\gg_0(\s)$ and to its
$sl(2,{\Bbb C})$ subalgebra  specified by $J_+ =  (\Lambda)_0$. 

We have defined two brackets $\{\cdot,\cdot\}^D$ and
$\{\cdot,\cdot\}^\ast$ in the set of functionals ${\rm Fun\/}(Q^{({\rm
can})})$, and the most important question is how to compare them.
In~[\Ref{WNOS},\Ref{GEN2}], it was shown that the second Poisson
bracket of the integrable hierarchies of~[\Ref{GEN1},\Ref{GEN2}] have 
non-trivial centres  that are in one-to-one correspondence with the elements of
the set ${\cal Z}$ defined in~\CentreZ. Nevertheless, the explicit
example presented in~[\Ref{WNOS}] shows that it is not generally true 
that all the components of $(q^{\rm can}(x))_1$ are centres of
$\{\cdot,\cdot\}^\ast$, which means that, in general, this bracket 
can be different to the Dirac bracket.

The defining property of $\{\cdot,\cdot\}^\ast$ is that the Poisson
algebras $({\rm Fun\/}_0(Q), \{\cdot,\cdot\}_2)$ and $({\rm
Fun\/}(Q^{\rm can}), \{\cdot,\cdot\}^\ast)$ are isomorphic. In
contrast, the Poisson structure induced by the Dirac bracket involves
the restriction to the invariant functionals with respect to the
transformations generated by the Hamiltonian vector fields associated
to some set of first-class constraints (see Section~3).
Taking into account this, we
can get another indication that the two Poisson brackets will be
generally different by analysing the infinitesimal gauge invariance
conditions satisfied by the elements of ${\rm Fun\/}_0(Q)$. The
condition associated to $S(x)\in C^\infty(S^1,P)$ is given by 
eq.~\GaugeInvar, and it can be expressed as
$$
0\>= \>\delta_S\varphi \> =\> \{ \varphi, \phi_S\}_2[q] \> + \> \Bigl( (d_q
\varphi)_{-1}\>, \> [S(x), \lambda_1 + (q(x))_1 ] \Bigr)\>,
\nfr{NoHam}
where $\phi_S[q] = \bigl(S(x), q(x) \bigr)$, 
which shows that all these constraints will be Hamiltonian only when the
choice of $\Lambda$, $\sw$ and $\s$ ensure that $Q\cap
\gg_1(\s)=\{0\}$, \ie, when all the components of $q(x)$ have zero
$\s$-grade.

The explicit relation between the two Poisson brackets
$\{\cdot,\cdot\}^D$ and $\{\cdot,\cdot\}^\ast$ can be obtained
by generalizing the proof of eq.~\Equal\ in Section~3. The starting
point will be the observation that, taking into account~\SplitFin, 
$\Delta^{i,j}(x,y)$, defined in eq.~\OpMat, has the block form
$$
\Delta^{i,j}(x,y)\> [q^{\rm can}] \> = \> \bordermatrix{
& \Gamma_0 & \Gamma_1 & \Gamma_2\cr
\Gamma_0 & 0 & 0 & A \cr
\Gamma_1 & 0 & B & C \cr
\Gamma_2 & D & E & F \cr} \>, 
\nfr{BlockDI}
which implies that
$$
\Delta_{i,j}(x,y)\> [q^{\rm can}] \> = \> \bordermatrix{
& \Gamma_0 & \Gamma_1 & \Gamma_2\cr
\Gamma_0 & \tilde F & \tilde E & D^{-1} \cr
\Gamma_1 & \tilde C & B^{-1} & 0 \cr
\Gamma_2 & A^{-1} & 0 & 0 \cr} \>,
\nfr{BlockDII}
where
$$
\eqalign{
\tilde C \> & = \> - B^{-1}\> C \> A^{-1} \cr
\tilde E \> & = \> - D^{-1} \>E \>B^{-1} \cr
\tilde F \> &= \> - D^{-1}\> F\> A^{-1} \> + \> D^{-1} \> E\> B^{-1}
\> C \> A^{-1}  \cr}
\nfr{FormInv}
are matrix differential operators. In the following, it will be important that 
$$
\eqalign{
B\>&=\> \langle [\theta^i,\theta^j]\>, \> J_+ +\left(q^{\rm
can}(x)\right)_0 \rangle_{\gg} \>\delta(x-y) \cr
& \equiv \> M^{i,j}[(q^{({\rm can})}(x))_0]\>\delta(x-y)\>,\cr}
\nfr{MatQ}
and, hence, that $M^{i,j}$ is just an antisymmetric
$(q^{({\rm can})}(x))_0$-dependent matrix for any $\theta^i,
\theta^j\in \Gamma_1$. 

Now, for any $\hat \varphi, \hat \psi\in C^\infty(S^1, Q^{\rm
can})$, let us consider the gauge invariant functions $\varphi[q] =
\pi^\ast(\hat \varphi)[q] = \hat\varphi[q^{\rm can}[q]]$ and
$\psi[q] = \pi^\ast(\hat \psi)[q] = \hat\psi[q^{\rm can}[q]]$, which,
obviously, satisfy eq.~\NoHam. Then
$$
\{ \varphi, \phi^i(x)\}_2[q] \>= \> - \Bigl\langle (d_q
\varphi)_{-1}\> ,\> [\theta^i, \lambda_1 +(q(x))_1] \Bigr\rangle_\gg\>,
\nfr{ConstA}
for any $\theta^i \in \Gamma_0\cup \Gamma_1 \subset P$, and, in
particular, the right hand side of~\ConstA\
vanishes when $\theta^i\in \Gamma_0$. Taking into account all this,
eq.~\DiracF\ becomes
$$
\eqalignno{
\noalign{\vskip 0.1cm}
\{\hat \varphi,  \hat \psi\}^\ast [q^{\rm can}] \> & =\>
\overline{\{ \varphi, \psi\}_2[q]} \cr 
\noalign{\vskip 0.1cm}
& =\> \{\hat \varphi, \hat \psi\}^D [q^{\rm can}] 
\> +\> {\cal C}(\hat \varphi,\hat \psi)\> [q^{\rm can}]\>, \qquad {\rm
with} \cr
\noalign{\vskip 0.3cm}
{\cal C}(\hat \varphi,\hat \psi)\>& [q^{\rm can}] \>  = \>
\sum_{\theta^i,\theta^j \in \Gamma_1} \> \int_{S^1} dx\> 
\Bigl\langle\> \overline{(d_q\varphi)_{-1}}\>, \>[\theta^i, \lambda_1 +
(q^{\rm can}(x))_{1}]\Bigr\rangle_\gg\cr
\noalign{\vskip 0.1cm}
&M_{i,j}[ (q^{\rm can} (x))_0]\> 
\Bigl\langle\> \overline{(d_q\psi)_{-1}}\>,
\>[\theta^j, \lambda_1 +
(q^{\rm can}(x))_1]\Bigr\rangle_\gg\>\>, & \nameali{GEND} \cr
\noalign{\vskip 0.1cm}
}
$$
where the antisymmetric matrix $M_{i,j}[(q^{\rm can}(x))_0]$ is
the inverse of $M^{i,j}[(q^{({\rm can})}(x))_0]$ (see eq.~\MatQ).

Since the Dirac bracket $\{\cdot,\cdot\}^D$ defines a
\W-algebra, eq.~\GEND\ explicitly shows that the second
Poisson bracket of the integrable hierarchies
of~[\Ref{GEN1},\Ref{GEN2}]  is a \W-algebra modified by the (polynomial) term
${\cal C}(\cdot, \cdot)$, which it is the main result of this paper. Because of
its importance, let us remind that we have restricted ourselves to those
integrable hierarchies where $\Lambda +q(x) \in \gg_0(\s)\oplus \gg_1(\s)$ and
$(\Lambda)_0 \not=0 $.

\section{The centres of $\{\cdot,\cdot\}^\ast$.}

The components of $q^{\rm can}(x)$ generate the
algebra defined by $\{\cdot,\cdot\}^\ast$, \ie, the
second Poisson bracket algebra of the integrable
hierarchy. In general, we can split them in two
sets. First, the components of  $(q^{\rm can}(x))_0$, which 
also generate the \W-algebra given by
the Dirac bracket $\{\cdot,\cdot\}^D$. The second set consists of the
components of  $(q^{\rm can}(x))_1$, which are centres
of the Dirac bracket, but, in general, not of $\{\cdot,\cdot\}^\ast$. 

The centres of $\{\cdot,\cdot\}^\ast$ are
in one-to-one relation with the elements of the set $\cal Z$ defined in
eq.~\CentreZ. Since they have vanishing brackets with all the
other generators of the second Poisson bracket algebra, they can be chosen
to be zero. This is equivalent to an additional trivial reduction, 
which means that the resulting subset of
generators form a Poisson subalgebra. Then, taking into account our choice of
$Q^{\rm can}$, we conclude that  the number of non-trivial generators
of the second Poisson bracket algebra equals the number of generators
of the associated \W-algebra plus 
$$
{\rm dim\/}(Q\cap \gg_1(\s))\> -\> {\rm dim\/}(P^\ast) \> -\> {\rm
dim\/}({\cal Z})\>.
\nfr{Dim}

The constraints that the centres of $\{\cdot, \cdot\}^\ast$ have to be
zero are equivalent to impose relations among the components of
$q^{\rm can}(x)$ that allow one to express some of them as polynomial
functions of the other ones. To be more precise, let us recall the
definition of the centres~[\Ref{WNOS},\Ref{GEN1},\Ref{GEN2}]. 
Consider the transformation
$$
\eqalign{
{\cal L}\> &= \> \exp\left( \ad V\right)\> (L) \>= \> L \> + \> [V,L]
\> + \> {1\over 2} [V,[V, L]]\> + \> \cdots \cr
& = \> \partial_x + \Lambda \> + \> h(x)\cr}
\nfr{Diagonal}
that ``abelianizes'' the Lax operator, where $V(x)\in C^\infty\bigl(S^1, {\rm
Im\/}(\ad \Lambda)\cap 
\gg_{<0}(\sw)\bigr)$ and $h(x) \in C^\infty\bigl(S^1, {\rm Ker\/}(\ad \Lambda)
\cap \gg_{<i}\bigr)$ are polynomial functionals
of the components of $q(x)$ and their $x$-derivatives.  
Then, when $i>1$, the functionals of the form $\Theta_b(x) = 
\overline{\langle b, h(x)\rangle}_{\gg}$, for any
$$
b\in {\cal Z}^\vee\>= \> \biggl[ {\rm Ker\/}(\ad \Lambda)\cap
\gg_{1-i}(\sw)\biggr]\> \cup \> \biggl[ {\rm Cent\/}\bigl({\rm
Ker\/}(\ad \Lambda)\bigr) \cap \Bigl[ \bigoplus_{j=1-i}^{-1} \gg_j(\sw)
\Bigr]\biggr]\>,
\efr
are gauge invariant centres of $\{\cdot, \cdot\}^\ast$; \ie, the
$\Theta_b$'s are the components of $h(x)$ along $\cal Z$. 
Moreover, when $b\in {\cal Z}^\vee \cap\gg_{-j}(\sw)$, it is 
quite easy to check that $\Theta_b(x)$ is a polynomial functional
only of the components of $q(x)$ whose $\sw$ grade is $\geq j$, and
that it is linear in the components of $q(x)$ whose $\sw$-grade equals
$j$. From
now on, we will use the notation $\Theta_j(x) = \Theta_b(x)$ when $b\in
{\cal Z}^\vee \cap\gg_{-j}(\sw)$, even though one should introduce an
additional index to indicate that ${\cal Z}^\vee$ could have more than
one linearly independent element in $\gg_{-j}(\sw)\>\>$\note{In addition to
these centres, the functionals $\Theta_b(x)$ where either $b\in {\rm
Cent\/}\bigl({\rm Ker\/}(\ad \Lambda)\bigr) \cap \gg_0(\sw)$, if $i>1$, or
$b\in {\rm Ker\/}(\ad \Lambda) \cap \gg_0(\sw)$, if $i=1$, are constant along
all the flows of the integrable hierarchy~[\Ref{GEN1}]. In the
second reference of~[\Ref{FHM}] and in~[\Ref{ARA}], these constant
functionals are related to the existence of ``residual'' gauge symmetries,
which is used there to induce an additional reduction of the second Poisson
bracket algebra. However, this additional reduction cannot be done by following
the Drinfel'd-Sokolov procedure and, therefore, the result is a non-polynomial
algebra.}.

Since the centres are gauge invariant functionals, one can construct
them directly in terms of the gauge-fixed Lax operator $L^{\rm
can} = \partial_x + \Lambda + q^{\rm can}(x)$; then, 
$$
\Theta_j(x) \> =\> \overline{\langle b\>, \> h(x) \rangle}_\gg \> = \>
\langle b\>, \> q^{\rm can}(x) \rangle_\gg \> + \> \cdots\>,
\nfr{Despeja}
where $\cdots$ indicate non-linear terms that are polynomial in the
components of $q^{\rm can}(x)$ whose $\sw$-grade is $>j$. 

Let us now show that $\langle b\>, \> q^{\rm can}(x) \rangle_\gg
\not=0$. Notice that ${\cal Z}\subset Q$, which, since the
bilinear form is non-degenerate, implies that $b$
cannot be orthogonal to all the elements of $Q$. Moreover, since ${\cal Z}^\vee
\subset {\rm Ker\/}(\ad \Lambda)$, the invariance of the bilinear form
also implies that $b$ is orthogonal to all the elements of $[\Lambda, P]$, and,
according to~\DSGF, we conclude that $\langle
b\>,  \> q^{\rm can}(x) \rangle_\gg \not=0$. 

Even more, whenever $(b_{-1})\not=0$, one can also
prove that $\langle (b)_{-1}, (q^{\rm can}(x))_1\rangle_\gg \not=0$. Notice
that $b\in {\cal Z}^\vee \subset
\Ker$; then, $[b, \Lambda]=0$, which implies that
$[(b)_0, J_+] + [b_{-1}, \lambda_1]=0$. Again, let us consider
eq.~\DSGF\ and our particular choice of gauge slice:
$$
Q \cap \gg_{1}(\s) \> = \> [\lambda_1\> , \> P^\ast] \> + \> 
Q_{\rm can} \cap \gg_{1}(\s)\>.
\efr
Since the bilinear form is non-degenerate, $(b)_{-1}\not=0$
cannot be orthogonal to all the elements of $Q\cap \gg_{1}(\s)$. But,
using that the bilinear form is invariant and that $P^\ast \subset
{\rm Ker\/}(\ad J_+)$, the identity $[b_{-1}, \lambda_1] = -
[(b)_0, J_+]$ implies that $(b)_{-1}$ is orthogonal to all the
elements of $[\lambda_1, P^\ast]$ and, therefore, that $\langle (b)_{-1},
(q^{\rm can}(x))_1\rangle_\gg \not=0$

Consequently, when the centres are taken to be zero, eq.~\Despeja\ can
be iteratively used,  
in order of descending $\sw$-grades starting with $\Theta_{i-1}$, to
eliminate all the generators of the second Poisson bracket algebra 
of the form $\langle b\>,  \> 
q^{\rm can}(x) \rangle_\gg$, for any $b\in{\cal Z}^\vee$. 

A logical question is whether this procedure can always be used to express 
certain components of $(q^{\rm can}(x))_1$ in terms of the components of 
$(q^{\rm can}(x))_0$. Since $Q\subset \gg_0(\s) \oplus \gg_1(\s)$, any
$b\in {\cal Z}^\vee$ has the decomposition $b= (b)_0 + (b)_{-1}$, and,
then,
$$
\langle b\>,  \>  q^{\rm can}(x) \rangle_\gg\> = \> 
\langle (b)_0\>,  \> (q^{\rm can}(x))_0 \rangle_\gg\> + \>
\langle (b)_{-1}\>,  \> (q^{\rm can}(x))_1 \rangle_\gg\>.
$$
Therefore, when the centre $\Theta_j(x)$ is associated to an element $b$
such that $(b)_{-1}\not= 0$, it can indeed be used to eliminate that 
component as
$$
\langle (b)_{-1}\>,  \> (q^{\rm can}(x))_1 \rangle_\gg \>= \>
-\> \langle (b)_{0}\>,  \> (q^{\rm can}(x))_0 \rangle_\gg \> + \>
\cdots\>, 
\nfr{Elimina}
where $\cdots$ indicate non-linear polynomial functionals of the
components of $q^{\rm can}(x)$ whose $\sw$-grade is $>j$.

On the contrary, whenever $(b)_{-1} = 0$, the condition $\Theta_j(x)=0$
allows one to eliminate the
component $\langle (b)_{0}\>,  \> (q^{\rm can}(x))_0
\rangle_\gg$, which is one of the generators of the \W-algebra given
by the corresponding Dirac bracket. This possibility only occurs if
$\Lambda$ is non-regular because, only in this case, ${\rm Ker\/}(\ad
\Lambda)\cap \gg_{i-1}(\sw)$ can contain nilpotent elements.

\chapter{Examples.}

In order to clarify the structure of the second Poisson bracket algebra, it will
be convenient to distinguish the two sets of generators $W_a(y)$ and $B_a(y)$
associated to the components of $q^{\rm can}[q(x)]$ whose
$\s$-grade equals $0$ and $1$, respectively. Then, for any
$\omega_a \in Q^{\rm can}\cap \gg_0(\s) = {\rm Ker\/}(\ad J_-)$
and $\beta_a \in Q^{\rm can}\cap \gg_{-1}(\s)$, we define the gauge
invariant functionals
$$
W_a(y) \> = \> \Bigl\langle \omega_a\>, \> q^{({\rm
can})}[q(y)]\Bigr\rangle_{\gg} \> ,
\nfr{WGen}
and
$$
B_a(y) \> = \> \Bigl\langle \beta_a\>, \> q^{({\rm
can})}[q(y)]\Bigr\rangle_{\gg} \> ;
\nfr{WGen}
we will also use $\hat{W}_a(y) = \overline{W_a(y)}$ and 
$\hat{B}_a(y) = \overline{B_a(y)}$. 

This way, according to~\GEND, the second Poisson bracket algebra can be
expressed as
$$
\eqalign{
& \{\hat{W}_a(y), \hat{W}_b(z)\}^\ast\> = \> \{\hat{W}_a(y),
\hat{W}_b(z)\}^D \> + \> {\cal C}\left(\hat{W}_a(y), \hat{W}_b(z)\right) \cr
& \{\hat{W}_a(y), \hat{B}_b(z)\}^\ast\> = \> {\cal C}\left(\hat{W}_a(y),
\hat{B}_b(z)\right) \cr
& \{\hat{B}_a(y), \hat{B}_b(z)\}^\ast\> = \> {\cal C}\left(\hat{B}_a(y),
\hat{B}_b(z)\right) \>, \cr}
\nfr{ExpForm}
where 
$$
\{\hat{W}_a(y), \hat{W}_b(z)\}^D \> =\> \sum_j
P^{j}_{a,b}(\hat{W}_1(y) , \hat{W}_2(y),\ldots)
\partial_{y}^{(j)}(y-z)
\efr 
is the \W-algebra associated to the $sl(2,{\Bbb C})$ subalgebra of
$\gg_0(\s)$ specified by $J_+=(\Lambda)_0$, and $P^{j}_{a,b}$ is a
differential polynomial. 

In~\GEND, ${\cal C}(\hat \varphi, \hat\psi)$ arises as a consequence of
the existence of non-vanishing components of $q(x)$ whose $\s$-grade
equals $1$. Actually, the factor ${\cal C}\left(\hat{W}_a(y),
\hat{W}_b(z)\right)$ is particularly interesting since it corresponds
to a possible deformation of the \W-algebra that would be induced by the
dependence of $(q^{\rm can} [q(x)])_0$ on the components of $(q(x))_1$.
According to eq.~\Orbit, $q^{\rm can}  [q(x)]$ is given by  
$$ 
\eqalign{
(q^{\rm can}[q(x)])_0\> & = \> \exp\left( \ad S^{({\rm
can})}[q(x)]\right) \Bigl( \partial_x \> + \> (\Lambda)_0 \> + \>
(q(x))_0 \Bigr)\cr
& \qquad\qquad  -\> \partial_x \>- \> (\Lambda)_0\cr
(q^{\rm can}[q(x)])_1\> & = \> \exp\left( \ad S^{({\rm
can})}[q(x)]\right) \Bigl( (\Lambda)_1 \> + \>
(q(x))_1 \Bigr) \> -\> (\Lambda)_1 \>.\cr}
\nfr{CompCan}
which makes clear that $(q^{\rm can} [q(x)])_0$ can only depend on
$(q(x))_1$ through $S^{\rm can}[q(x)]$. Now, considering the
choice of  $Q^{\rm can}$ made in Section~5, it follows
that $S^{\rm can}[q(x)]$ depends on $(q(x))_1$ only when $P^\ast =
{\rm Ker\/}(\ad J_+) \cap P\not = \{0\}$. 

Therefore, ${\cal C}(\hat{W}_a(y), \hat{W}_b(z))$ vanishes when
$P^\ast = \{0\}$, and, in this case, the
restriction of $\{\cdot, \cdot\}^\ast$ to the $\hat{W}_a$'s is just
the \W-algebra given by the Dirac bracket. $P^\ast = \{0\}$ is equivalent
to the non-degeneracy condition~\DegenPlus, and
this is precisely the case discussed in~[\Ref{WNOS}]. 
Now, using eq.~\GEND, we can clarify the results of that paper, and this will
constitute our first example. 

Nevertheless, the main motivation of this work was to investigate the form of
the second Poisson bracket algebras precisely when the
non-degeneracy condition is not satisfied, which is illustrated by the other
examples. They correspond to some integrable hierarchies
associated to $A_{N-1}$ and to the following data:
$[w]=[N]$ is the conjugacy class of the Coxeter element, which means
that ${\cal H}[w]={\cal H}[N]$ is the principal Heisenberg subalgebra,
$\sw=(1,1,\ldots,1)$ is the principal gradation, $\s=(1,0,\ldots,0)$
is the homogeneous gradation, and $\Lambda\in {\cal H}[N]$ has
grade $1<i <N$. Since these hierarchies have flow equations with fractional
scaling dimension, and following the terminology of~[\Ref{BAK}], we will refer
to them as fractional $A_{N-1}$ KdV hierarchies, and we will use the
notation $[N]^i$. 

Even though in all these examples the non-degeneracy condition is not
satisfied, their second Poisson bracket algebra is given just by the Dirac
bracket, which means that all the components of
$(q^{\rm can}(x))_1$ can be expressed as functionals of the components of
$(q^{\rm can}(x))_0$, and that ${\cal C}(\cdot,\cdot)$ vanishes. 
We present them here because they clarify the results obtained by other
authors. However, it
is clear that they can hardly be used to gain intuition about the general case.
The reason is that they exhibit very special features that should not be
expected to hold in general. For instance, consider the semisimple element
$J_0$ of the $sl(2,{\Bbb C})$ subalgebra of
$A_{N-1}$ specified by $J_+ = (\Lambda)_0$, which has been obtained
in~[\Ref{FREP},\Ref{COND}]; for $[N]^i$, if $N= mi +r$ with $m=[N/i]$
and $0\leq r<i$, the result is (see eq.~(3.68) of~[\Ref{FREP}])
$$
J_0\> = \> {\rm diag\/} \biggl(
\underbrace{{m/2},\ldots,{m/2}}_{r\;{\rm times}} \> ,\>
\underbrace{{(m-1)/2},\ldots,{(m-1)/2}}_{(i-r)\;{\rm times}}\>,\>
\ldots\>, \>
\underbrace{{-m/2},\ldots,{-m/2}}_{r\;{\rm times}}\biggr)\>,
\efr
which means that the gradation induced by $J_0$ on $\gg_0(\s)\simeq
A_{N-1}\otimes 1$ assigns non-negative grade to all the raising operators
$e_{1}^+, \ldots, e_{r}^+$ corresponding to the same basis
for the simple roots used in eq.~\Derivation. Then, the gradation induced by
$J_0$ is comparable with the gradation $\sw$ through the partial ordering
defined in Section~2 and, since the principal gradation is always maximal, it
follows that (see eq.~\Repr) 
$$
\eqalign{
&\gg_0(\sw) \subset 
\bigoplus_{k=1}^n \bigl\{ X_{0}^{(j_k)}(y_k) \bigr\}\>, \cr
&\bigoplus_{k=1}^n \bigl\{ X_{m}^{(j_k)}(y_k)\bigm| m>0\bigr\} \subset
\gg_0(\s) \cap \gg_{>0}(\sw) \cr
&\bigoplus_{k=1}^n \bigl\{ X_{m}^{(j_k)}(y_k)\bigm| m<0\bigr\} \subset
\gg_0(\s) \cap \gg_{<0}(\sw) \cr}
\nfr{Include}
as a very particular characteristic of all the fractional KdV $A_{N-1}$
hierarchies. In~\Include, we have labelled the elements of the irreducible
representation $D_{j_k}(y_k)$ by the eigenvalue of $J_0$, \ie, $[J_0,
X_{m}^{(j_k)}(y_k) ]= m X_{m}^{(j_k)}(y_k)$. An straightforward
consequence of~\Include\ is that all the elements of $P^\ast = {\rm
Ker\/}(\ad J_+)\cap \gg_{<0}(\sw)$ are spin-singlets, which 
enables a very convenient simplification of the
gauge fixing procedure that will be illustrated by our last example.

Finally, let us also point out another two important particular features
exhibited by our examples. First, the number of centres equals the dimension of
$Q^{\rm can}\cap \gg_1(\s)$, and, since $\Lambda$ is regular, all the
components of $(q^{\rm can}(x))_1$ can be expressed as polynomial
functionals of the components of $(q^{\rm can}(x))_0$. The second is
that $\overline{P}= \Gamma_0 \cup \Gamma_1$ is a subalgebra of $P$ (see
eq.~\PPartA), which, in eq.~\MatQ, implies that $\langle [\theta^i,
\theta^j], (q^{\rm can}(x))_0\rangle_{\gg}=0$ and, hence,
$M_{i,j}[(q^{\rm can}(x))_0]$ is just a constant matrix.

Going beyond these particular cases would require a detailed analysis of
eqs.~\GEND\ and \ExpForm. In particular, it would be important to know if there
exists an energy-momentum tensor that generates the conformal symmetry of the
second Poisson bracket algebra. If this happens together with a
non-trivial ${\cal C}(\cdot, \cdot)$, it might lead to the construction of
new extended conformal algebras different that those associated to the
$sl(2,{\Bbb C})$ subalgebras of a Lie algebra through Drinfel'd-Sokolov
Hamiltonian reduction. Regretfully, those very interesting
cases must involve large rank Lie algebras, and we have not yet  been able
to find any example. 

\section{The second Poisson bracket algebra when ${\rm Ker\/}(\ad
(\Lambda)_0) \cap P =\{0\}$.}

Let us start by studying the form of the matrix in eq.~\MatQ. Since the
bilinear form of $\gg$ provides a one-to-one map between ${\rm
Ker\/}(\ad J_+)$ and $\gg_{<0}(\sw)$ and
${\rm Ker\/}(\ad J_-)$ and $\gg_{>0}(\sw)$, respectively, the
non-degeneracy condition~\DegenPlus\ is equivalent to
$$
{\rm Ker\/}(\ad J_-) \cap \Bigl[ \gg_0(\s) \cap \gg_{>0}(\sw)\Bigr]
\> =\> \{0\}\>.
\nfr{DegenPB}
Then, with our choice of $Q^{\rm can}$, it follows that $Q^{\rm can}\cap
\gg_0(\s) \subset \gg_{\leq0}(\sw)$, and, hence, eq.~\MatQ\ simplifies to
$$
M^{i,j}[(q^{\rm can}(x))_0]\> =\> \bigl\langle [\theta^i,
\theta^j]\>, \> J_+\bigr\rangle_{\gg}\>,
\efr
which shows that $M_{i,j}$ is simply a constant antisymmetric matrix.

The condition $P^\ast=\{0\}$ ensures that $S^{\rm can}[q(x)]$ is
independent of
$(q(x))_1$, and, hence, that $(d_q W_a(y))_{-1} = 0$. Moreover, it also
allows the calculation of $(d_q  B_a(y))_{-1}$. 
According to eq.~\CompCan, $B_a(y)$ is given by
$$
B_a(y) \>= \>  \Bigl\langle \beta_a\>, \> (q(y))_1 \> +\>
\bigl[S^{\rm can}[(q(x))_0]\>, \> (q(x))_1\bigr]\> + \> \cdots
\Bigr\rangle_{\gg} \>.
\nfr{BaExp}
Then, using that $S^{\rm can}[q(x)]$ vanishes when $q(x) \in
C^\infty(S^1, Q^{\rm can})$, and that $S^{\rm can}[q(x)]$ is
independent of $(q(x))_1$, it follows that 
$$
\overline{(d_q B_a(y))_{-1}}\> =\> \beta_a \> \delta(x-y)\>.
\nfr{DerB}

All this shows that, when the non-degeneracy condition~\DegenPlus\ 
is satisfied, the second Poisson bracket algebra~\ExpForm\ becomes
$$
\eqalignno{
& \{\hat{W}_a(y), \hat{W}_b(z)\}^\ast\> = \> \{\hat{W}_a(y),
\hat{W}_b(z)\}^D \cr
& \{\hat{W}_a(y), \hat{B}_b(z)\}^\ast\> = \> 0 \cr
& \{\hat{B}_a(y), \hat{B}_b(z)\}^\ast\> = \> 
-\sum_{\theta^i, \theta^j \in \Gamma_1} \Bigl\langle \beta_a\>, \>
[\theta^i, \lambda_1 + (q^{\rm can}(y))_1] \Bigr\rangle_{\gg}\>
M_{i,j}\cr
\noalign{\vskip 0.1cm} 
&\qquad\qquad\qquad  \Bigl\langle \beta_b\>, \>
[\theta^i, \lambda_1 + (q^{\rm can}(y))_1] \Bigr\rangle_{\gg} \>
\delta(y-z)\>, &\nameali{ExpFormP} \cr}
$$
which shows that the $\hat{W}_a(y)$'s generate the \W-algebra
corresponding to the Dirac bracket, and that the two sets of generators
$\hat{W}_a(y)$ and $\hat{B}_b(y)$ are actually decoupled.

The explicit form of the second Poisson bracket given by eq.~\ExpFormP\
allows one to investigate the existence of an energy-momentum tensor in this
case. Since the two sets of generators $\hat{W}_a(y)$ and $\hat{B}_b(y)$ are
decoupled, it has to be of the form $T(x) = {\cal T}(x) + \Delta T(x)$,
where ${\cal T}(x)$ is the energy-momentum tensor of the \W-algebra
generated by the $\hat{W}_a(y)$'s and $\Delta T(x)$ generates the conformal
transformation of the $\hat{B}_b(y)$'s. Since $\Delta T(x)$ has to be a
differential polynomial functional of $\hat{B}_b(y)$, the form of
$\{\hat{B}_a(y), \hat{B}_b(z)\}^\ast$ shows that it will be impossible to
find $\Delta T(x)$ unless all the terms that depend on $(q^{\rm can}(x))_1$
in~\ExpFormP\ vanish, which, for instance, happens if the $\sw$-grade of
$\Lambda$ is $i=2$. In such case, the relevant bracket simplifies to
$$
\eqalign{
\{\hat{B}_a(y), \hat{B}_b(z)\}^\ast\> &= \> 
-\sum_{\theta^i, \theta^j \in \Gamma_1} \bigl\langle \beta_a\>, \>
[\theta^i, \lambda_1] \bigr\rangle_{\gg}\>
M_{i,j}\> \bigl\langle \beta_b\>, \>
[\theta^i, \lambda_1 ] \bigr\rangle_{\gg} \>
\delta(y-z)\cr
& \equiv \> \Omega_{a,b}\>\delta(y-z)\>, \cr}
\efr
where $\Omega_{a,b}$ is an antisymmetric non-degenerate matrix, if we assume
that all the centres of the second Poisson bracket have already been removed.
Since $\Omega$ is non-degenerate, it is possible to choose a basis
$\hat{B}_{b}^\pm(y)$ such that the only non-vanishing Poisson brackets are
$$
\{\hat{B}_{a}^+(y), \hat{B}_{b}^-(z)\}^\ast\> = \> \delta_{a,b}\>
\delta(y-z)\>,
\efr
which means that the restriction of the second Poisson bracket algebra to the
$\hat{B}_b(y)$'s is just a set of decoupled ``b--c'' algebras. Therefore, if
the total number of generators is $2N$, the corresponding energy-momentum
tensor is just
$$
\Delta T(x) \> =\> - \sum_{a=1}^N \biggl[ \Delta_a \>
\bigl(\hat{B}_{a}^+(x)\bigr)'\> \hat{B}_{a}^-(x) \> +\>
(\Delta_a -1)\> \hat{B}_{a}^+(x)\> \bigl(\hat{B}_{a}^-(x)\bigr)'
\biggr]\>,
\efr
and it assigns conformal dimensions $\Delta_a$ and $1-\Delta_a$ to the
generators $\hat{B}_{a}^+(x)$ and $\hat{B}_{a}^-(x)$, respectively, where
the $\Delta_a$'s are completely arbitrary real numbers. As a particular example
of this construction, let us refer to the KdV-hierarchy associated to ${\cal
H\/}[3,3]\subset A_{5}^{(1)}$ discussed in~[\Ref{WNOS}]. 

\section{The fractional $[N]^3$ hierarchies.}

The fractional $[N]^2$ hierarchies
have already been discussed in~[\Ref{WNOS}]. In these cases, 
$\Lambda$ satisfies the condition~\Degen\ only when $N$
is odd, and, then, it also satisfies the non-degeneracy
condition~\DegenPlus~[\Ref{COND}]. Therefore, its second Poisson
bracket algebra is the \W-algebra associated to the $sl(2,{\Bbb C})$
subalgebra labelled by the partition
$$
N\> =\> \left(N+1\over 2\right) \> +\> \left(N-1 \over 2\right)\>,
\efr
which is nothing else than the fractional ${\cal W}^{(2)}_N$ algebra
of~[\Ref{COND},\Ref{POLY}].

Let us now consider the fractional $[N]^3$ hierarchies,
where\note{From 
now on, we will use the defining representation of $A_{N-1}$ in
terms of traceless $N\times N$ matrices, and ${\Bbb I}_k$ will be the
$k\times k$ identity matrix.} 
$$
\Lambda \>= \> \pmatrix{ 0 & {\Bbb I}_{N-3} \cr
z {\Bbb I}_{3}& 0\cr} \>, \qquad N\geq4\>,
\efr
which satisfies the
condition~\Degen\ only if $N \not\in 3{\Bbb Z}$~[\Ref{WNOS}]; therefore,
we restrict ourselves to this case when, moreover,
$\Lambda$ is regular, \ie, $\Ker = {\cal H\/}[N]$. In contrast, 
$(\Lambda)_0$ does not satisfy the condition~\DegenPlus, and $P^\ast$
is the one-dimensional subspace generated by~[\Ref{WNOS}]
$$
\sum_{j=1}^{(N-1)/3} E_{3j, 3j-1} \>,
\efr
when $N\in 1+ 3{\Bbb Z}$, and by
$$
\sum_{j=1}^{(N+1)/3} E_{3j-1, 3j-2} \>,
\efr
when $N\in 2+ 3{\Bbb Z}$. 

Since $P^\ast$ is one-dimensional, $S^{\rm can}[q(x)]$ will be a functional of
the components of $(q(x))_0$ and of a single component of $(q(x))_1$. This means
that $\overline{(d_q W_a)_{-1}}$ is a function of $x$ taking values on certain
one-dimensional subspace of
$\gg_{-1}(\s)$.  Then, since $M_{i,j}[(q^{({\rm can})}(x))_0]$ is
antisymmetric, ${\cal C\/}\bigl(\hat W_a(y), \hat W_b(z)\bigr)$
vanishes and the restriction of the second Poisson bracket to the $\hat
W_a(y)$'s is given just by the Dirac bracket in this case. 

In particular, this means that all the components of $(q^{\rm can}(x))_1$ can
be eliminated by taking the centres to be zero, which can be explicitly
checked by using eq.~\Dim. Since $\Lambda$ is regular, ${\cal Z}$ is generated
by  the elements of the principal Heisenberg subalgebra whose principal grade
is $1$ and $2$,
\ie,
${\rm dim}({\cal Z}) = 2$. Moreover, ${\rm dim}(Q\cap
\gg_1(\s))= 3$, and ${\rm dim}(P^\ast)=1$.

Therefore, the second Poisson bracket algebra corresponding to these
fractional KdV hierarchies is the \W-algebra associated to the
$A_1 = sl(2,{\Bbb C})$ subalgebra of $A_{N-1}$ labelled by the
partition~[\Ref{WNOS}]
$$
N\> =\> \left(N+2 \over 3\right) \>+ \> \left( N-1\over 3\right) \> +
\> \left( N-1\over 3\right)\>,
\nfr{Tres}
when $N\in 1 + 3{\Bbb Z}$, and 
$$
N\> =\> \left(N+1 \over 3\right) \>+ \> \left( N+1\over 3\right) \> +
\> \left( N-2\over 3\right)\>,
\efr
when $N\in 2 + 3{\Bbb Z\>}$\note{After the work of~[\Ref{POLY}], the name
fractional \W-algebra, or ${\cal W}_{N}^l$
has been used in a quite confusing way to denote different extensions
of the conformal algebra obtained by reduction of the current algebra of
$A_{N-1}$. Since all the cases where the resulting algebra is
polynomial can be associated to some $sl(2, {\Bbb C})$ subalgebra of
$A_{N-1}$~[\Ref{COND}], we will use it to label the different
\W-algebras that are related to the fractional KdV
hierarchies. Nevertheless, let us notice that the \W-algebra
corresponding to $[4]^3$, which, according to~\Tres, is associated to the
partition $4=2+1+1$, is just the ``${\cal W}^{(3)}_4$'' 
algebra of~[\Ref{DMAT}].}.

According to~[\Ref{WNOS}], and using the same notation as in the Theorem~3
of that reference, the \W-algebra corresponding
to $[4]^3$ is also the second Poisson bracket
algebra of the integrable
hierarchy associated to the conjugacy class $[w]=[2,1,1]$ of the Weyl
group of $A_3$, with $\Lambda=\Lambda^{(1)}$, or to $[w]=[3,1]$, with
$\Lambda= \Lambda^{(2)}$. Also, the \W-algebra corresponding to $[5]^3$
is the second Poisson bracket algebra of the 
integrable hierarchy constructed from $[w]=[2,2,1]$ and $\Lambda=
\Lambda^{(1)}$. In contrast, all the \W-algebras corresponding to
$[N]^3$ with $N\geq 7$ are not the second Poisson 
bracket algebra of any generalized
integrable hierarchy constructed from a $\Lambda$ that satisfies the
non-degeneracy condition~\DegenPlus\ (see Theorem~3 of~[\Ref{WNOS}]).

Finally, let us point out that, although these examples consider only some
particular cases involving the $A_n$ algebras, in general, 
they show that whenever the subspace $P^\ast$
is one-dimensional, the restriction of the second Poisson bracket to
the $\hat W_a(y)$'s will be given just by
the Dirac bracket and, hence, that it will be the \W-algebra
associated to the $sl(2,{\Bbb C})$ subalgebra of $g$ specified by
$J_+=(\Lambda)_0$. 

\section{The fractional $[5]^4$ hierarchy.}

In our last example, the non-degeneracy
condition~\DegenPlus\ is not satisfied either, but, now, 
${\rm dim\/}(P^\ast) >1$.

Let us consider the fractional $A_4$ hierarchy corresponding to the
principal Heisenberg subalgebra and to the element 
$$
\Lambda \> =\> \pmatrix{ & & & & 1\cr z & & & & \cr & z& & & \cr
& & z& & \cr & & & z& \cr}
$$
with principal grade $i=4$, which is regular. Then, $Q$ is of the form
$$
\pmatrix{\ast & \ast & \ast & \ast & \cr
\ast &\ast &\ast &\ast &\ast &\cr
\ast &\ast &\ast &\ast &\ast &\cr
\ast &\ast &\ast &\ast &\ast &\cr
\ast &\ast &\ast &\ast &\ast &\cr} \> +\> 
z\> \pmatrix{&&&&\cr &&&&\cr \ast &&&&\cr
\ast& \ast& &&\cr \ast&\ast&\ast&&\cr}\>,
\efr   
while $P$ is the set of lower triangular matrices. 

In this case, $(\Lambda)_0$ specifies the $sl(2,{\Bbb C})$ subalgebra
of $A_5$ generated by
$$
J_+\> =\> (\Lambda)_0 \> = \> E_{1,5}\>, \quad
J_-\> = \> {1\over2}E_{5,1}\>, \quad {\rm and} \quad
J_0 \>=\> {1\over2}\left(E_{1,1}- E_{5,5}\right)\>,
\efr
which is labelled by the partition $5=2+1+1+1$, and, 
according to Section~5,
$$
Q^{\rm can}\cap \gg_0(\s)\> = \> {\rm Ker\/}(\ad J_-) \>= \> 
\pmatrix{a&&&&\cr
\ast&\ast &\ast &\ast &\cr \ast &\ast &\ast &\ast &\cr
\ast &\ast &\ast &\ast &\cr \ast &\ast &\ast &\ast &a\cr}\>,
\nfr{ExaForm}
and $P^\ast\>= \> P\cap {\rm Ker\/}(\ad J_+)$ is generated by the
three elements $E_{3,2}$, $E_{4,3}$, and $E_{4,2}$.

Let us consider eq.~\Dim. Since
$\Lambda$ is regular, $\cal Z$ is generated by the elements
of the principal Heisenberg subalgebra
of $A_{4}^{(1)}$ whose principal grade is $1$, $2$, and $3$; hence, 
${\rm dim\/}({\cal Z}) = 3$. Moreover, ${\rm
dim\/}(Q\cap \gg_1(\s)) = 6$, and ${\rm dim\/}(P^\ast)= 3$. Therefore,
since $b_{-1}\not=0$ for any $b\in{\cal Z}^\vee$, eq.~\Elimina\ ensures
that all
the components of $(q^{\rm can}(x))_1$ can be expressed as polynomial
functionals of the components of $(q^{\rm can}(x))_0$ when the centres
are chosen to be zero.

In order to use eq.~\GEND, we need
$$
\Gamma_1 \>  = \> {\rm Im\/}(\ad J_-) \cap \bigl[\gg_{>-4}\cap
\gg_{<0}\bigr] \>
= \> \pmatrix{&&&&\cr \ast&&&& \cr\ast&&&& \cr\ast&&&& \cr
&\ast&\ast&\ast& \cr} \>,
\efr
for which we choose the following basis 
$$
\{\theta^1, \ldots, \theta^6\} \>=\> \{E_{2,1},E_{3,1}, E_{4,1},
E_{5,2}, E_{5,3}, E_{5,4}\} \>.
\nfr{BasF}
Then,
$$
[\theta^j , \theta^k]\>= \> \bigl(\delta_{j, k+3} \> -\>
\delta_{j+3,k}\bigr)\> E_{5,1}
\nfr{Fun}
is orthogonal to all the elements of $Q^{\rm can}$, and, therefore,
according to~\MatQ, the matrix
$$
M_{j,k}[(q^{\rm can}(x))_0] \>= M_{j,k} \> =\> \delta_{j+3,k} \>- \>
\delta_{j,k+3} 
\efr
is just a constant matrix as expected; eq.~\Fun\ ensures that, also in this
case, $\overline{P}$ is a subalgebra. Consequently, the additional term
in~\GEND\ becomes
$$
\eqalignno{
{\cal C\/}(\hat \varphi, \hat\psi) & [q^{\rm can}]\> = \>
\sum_{j=1}^3 \int_{S^1} dx \biggl( \bigl\langle
\overline{(d_q\varphi)_{-1}}\>, \> [\theta^j, \lambda_1 + (q^{\rm
can}(x))_1] \bigr\rangle_{\gg} \cr
\noalign{\vskip0.3cm}
&\bigl\langle
\overline{(d_q\psi)_{-1}}\>, \> [\theta^{j+3}, \lambda_1 + (q^{\rm
can}(x))_1]\bigr\rangle_{\gg}\>
-\> (\varphi \leftrightarrow \psi) \biggr)\>. &\nameali{ExHard}}
$$

So far, we have only specified the choice of $Q^{\rm can}\cap
\gg_0(\s)$, but, to calculate the additional term~\ExHard, we also
need $Q^{\rm can}\cap \gg_1(\s)$ in order to obtain the dependence of
$S^{\rm can}$ in the components of $(q(x))_1$, and the form of 
$(q^{\rm can}(x))_1$. In this particular
example this can be easily done because, as a particular feature that
we have already announced, 
$$
P^{\ast}\> \subset \> {\rm Ker\/}(\ad J_-)\>.
\nfr{Feat}
Then, since ${\rm Ker\/}(\ad J_-)$ is a subalgebra of $\gg_0$, the
$q(x)$-dependent gauge transformation that takes $q(x)$ to $q^{\rm
can}[q(x)]$ can be constructed in the following simple way. First,
let us
consider a gauge transformation generated by $\overline{S}(x)\in
C^{\infty}(S^1, \overline{P})$ such that
$$
(\overline{q}(x))_0 \> =\> \exp \bigl(\ad \overline{S}(x)\bigr)
\Bigl(\partial_x \> + \> J_+ +\> (q(x))_0 \Bigr) \> -\> \partial_x \>
- \> J_+ \> 
\efr
is an element of $C^\infty\bigl(S^1, {\rm Ker\/}(\ad J_-)\bigr)$;
obviously, $\overline{S}(x)$ and $(\overline{q}(x))_0$ are local
functionals only of $(q(x))_0$, and, correspondingly, 
$$
(\overline{q}(x))_1 \> =\> \exp \bigl(\ad \overline{S}(x)\bigr)
\Bigl(\lambda_1 \> +\> (q(x))_1 \Bigr) \> -\> \lambda_1\>.
\efr
The second step is to specify the gauge transformation generated by
$S^\ast(x) \in C^{\infty}(S^1, P^\ast)$ that fixes the form of $Q^{\rm
can}\cap \gg_1(\s)$:
$$
(q^{\rm can}(x))_1\> =\> \exp \bigl(\ad S^\ast(x)\bigr)
\Bigl(\lambda_1 \> +\> (\overline{q}(x))_1 \Bigr) \> -\>
\lambda_1\>,
\nfr{Eone}
which makes sense because eq.~\Feat\ ensures that
$$
(q^{\rm can}(x))_0\> =\> \exp \bigl(\ad S^\ast(x)\bigr) \Bigl(
\partial_x \> + \> J_+ +\> (\overline{q}(x))_0 \Bigr) \> -\> \partial_x \>
- \> J_+ 
\nfr{Etwo}
is in $C^\infty\bigl(S^1, {\rm Ker\/}(\ad J_-)\bigr)$
too. Therefore, $S^{\rm can}[q(x)]$ is given by
$$
\exp\left( \ad S^{\rm can}[q(x)]\right)\> = \>
\exp\left(\ad S^\ast[q(x)]\right) \circ
\exp\left(\ad \overline{S}[(q(x))_0]\right)\>,
\efr
and one has to remember that 
$\overline{S}[(q(x))_0]$, $S^\ast[q(x)]$, and $S^{\rm can}[q(x)]$
vanish for $q(x) \in C^{\infty}(S^1,Q^{\rm can})$.

Since we only need the dependence on $(q(x))_1$, we
only have to obtain $S^\ast(x)$, which can be written as
$$
S^\ast(x)\>= \> \alpha(x) \> E_{3,2} \> + \> \beta(x) \> E_{4,3} \>+\>
\gamma(x) \> E_{4,2}\>;
\efr
recall that $\alpha(x)$, $\beta(x)$, and $\gamma(x)$ will be local
functionals of $(q(x))_0$ and $(q(x))_1$ that vanish when $q(x)\in
C^\infty(S^1, Q^{\rm can})$. According to~\ExaForm, $(q(x))_1$ can be
written as 
$$
(q(x))_1 \> = \> z\> \pmatrix{&&&&\cr &&&&\cr a(x) &&&&\cr
d(x)&b(x)&&&\cr f(x)&e(x)& c(x)&&\cr}\>;
\efr
then, eqs.~\Eone\ and \Etwo\ become
$$
\eqalignno{
a^{\rm can}(x) \> &= \> \overline{a}(x) \> + \> \alpha(x) \> =\> a(x)
\> + \> \alpha(x) \> + \> \cdots\cr
b^{\rm can}(x) \> &= \> \overline{b}(x) \> + \> \beta(x) \>-\>
\alpha(x) \>  =\> b(x)
\> + \> \beta(x) \>-\> \alpha(x) \> + \> \cdots\cr
c^{\rm can}(x) \> &= \> \overline{c}(x) \> - \> \beta(x) \> =\> c(x)
\> - \> \beta(x) \> + \> \cdots\cr
\noalign{\vskip 0.3cm}
d^{\rm can}(x) \>& =\> \overline{d}(x) \> +\> \gamma(x) \> + \>
\overline{a}(x) \> \beta(x) \> + \> {1\over2}\alpha(x) \> \beta(x) \>
= \> d(x) \> + \> \gamma(x) \> + \> \cdots\cr
e^{\rm can}(x) \>& =\> \overline{e}(x) \> -\> \gamma(x) \> - \>
\overline{c}(x) \> \alpha(x) \> + \> {1\over2}\alpha(x) \> \beta(x) \>
= \> e(x) \> - \> \gamma(x) \> + \> \cdots\cr
\noalign{\vskip 0.3cm}
f^{\rm can}(x)\> & =\> f(x)\> + \> \cdots \>,& \numali \cr}
$$
where the dots correspond to products of two or more components of
$\overline{S}(x)$ and $S^\ast(x)$ that vanish when restricted to the
gauge slice and, hence, they will not contribute to
$\overline{(d_q\varphi)_{-1}}$. Then, a possible choice of the gauge
slice, compatible with Section~5, is specified by
$$
Q^{\rm can}\cap \gg_1(\s) \> = \> z\>\pmatrix{&&&&\cr &&&& \cr
0&&&&\cr 0&0&&& \cr \ast&\ast&\ast&&\cr}\>,
\efr 
which means that
$$
\eqalign{
&a^{\rm can}(x)=0 \> \Rightarrow \alpha(x) \> = \> -\> a(x)\> + \> \cdots\cr 
&b^{\rm can}(x)=0 \> \Rightarrow \beta(x) \> = \> -\> a(x)\> -\> b(x)
\> + \>\cdots\cr 
&d^{\rm can}(x)=0 \> \Rightarrow \gamma(x) \> = \> -\> d(x)\> + \cdots \>,\cr}
\efr 
and that
$\overline{(d_q W_a)_{-1}}$ will be a function of $x$ taking values
in the subspace of $\gg_{-1}(\s)$ generated by $\{z^{-1}
E_{1,3}\>,\>\> z^{-1} E_{2,4}\>,\>\> z^{-1} E_{1,4}\}$. 

Consequently, using~\BasF\ and~\ExHard, one concludes that   
the additional term ${\cal
C}(\hat W_a(y), \hat W_b(z))$ vanishes identically, and that, again, 
the restriction of the second Poisson bracket to the $\hat W_a$'s is
just the \W-algebra associated to the $sl(2,{\Bbb C})$ subalgebra of
$A_4$ labelled by the partition $5=2+1+1+1$. Notice that, according
to~[\Ref{WNOS}], this integrable hierarchy has the same second Poisson
bracket algebra as the hierarchies associated to the conjugacy
classes $[w]=[2,1,1,1]$ and $[w]=[3,1,1]$ of the Weyl group of $A_4$, and to
$\Lambda=\Lambda^{(1)}$ and $\Lambda^{(2)}$, respectively.

Finally, let us point out that the previous analysis can be extended
for the generic fractional $[N]^{N-1}$ hierarchy, which shows that
its second Poisson bracket algebra is just the \W-algebra
associated to the partition $N= 2+ 1+ \cdots + 1$.

\chapter{Conclusions.}

The main result of this paper is summarized by eq.~\GEND, which
shows that the second Poisson bracket algebra of the integrable
hierarchies of~[\Ref{GEN1},\Ref{GEN2}] corresponds to a deformation of one of
the \W-algebras obtained by (classical) Drinfel'd-Sokolov reduction. Those
integrable hierarchies of partial differential equations are
associated to the non-conjugate Heisenberg subalgebras of the loop
algebra $\gg$ of a finite simple Lie algebra. Their construction
involves, first, the choice of a $\Bbb Z$-gradation $\sw$ of $\gg$ that is
compatible with the Heisenberg subalgebra; next, a second gradation
$\s$ that is $\preceq\sw$ with respect to a partial ordering,
and, finally, a constant graded element $\Lambda$ of
the Heisenberg subalgebra. In this paper, we have considered the
most general case that, according to~[\Ref{WNOS}],
is expected to be related to the \W-algebras
obtained by Drinfeld-Sokolov reduction. Thus, the \W-algebra that
is related to the second Poisson bracket algebra is specified by the zero
$\s$-graded (reductive) subalgebra of $\gg$ and by the $sl(2,{\Bbb C})$
subalgebra whose $J_+$ is the zero $\s$-graded component of $\Lambda$.

Although our results suggest that second Poisson bracket algebras could lead to
non-trivial deformations of the already known \W-algebras, and, maybe, to
different extended conformal algebras, we have not succeeded yet in finding an
example exhibiting that feature, but we find no reason to exclude that
possibility when considering the integrable hierarchies corresponding to large
rank  Lie algebras. 

Instead, eq.~\GEND\ allows one to characterize 
two general families of hierarchies where the deformation is trivial, \ie,
where the second Poisson bracket algebra consists of a \W-algebra plus some  
additional generators. 
The first family consists of those integrable hierarchies for which
the subalgebra $P^\ast = {\rm Ker\/}(\ad J_+) \cap \bigl[ \gg_0(\s)
\cap \gg_{<0}(\sw)\bigr]$ is empty, and, in this case, the additional
generators are not coupled to the \W-algebra. This is the only case actually
considered in~[\Ref{WNOS}], even though the decoupling was not shown
there. Notice that all the particular cases where $\Lambda$ is the element of
the Heisenberg subalgebra with minimum positive $\sw$-grade and this grade
equals $i=(\sw)_0 \geq1$  are included in this family; consequently, their
second Poisson bracket algebra is just the
\W-algebra associated to $J_+$. The second family includes all those
hierarchies for wich $P^\ast$ is one-dimensional, and their simple analysis
by means of eq.~\GEND\ is a nice example of how useful that equation is. In
relation to~[\Ref{WNOS}], this second family provide new examples of
\W-algebras that can be indentified with the second Poisson bracket algebra of
the integrable hierarchies of~[\Ref{GEN1},\Ref{GEN2}], even though we are
still far from proving that this relation works for all the \W-algebras
obtained through Hamiltonian reduction. Actually, it is known that the
second Poisson bracket algebra of the integrable hierarchies specified by the
non-degeneracy condition~\DegenPlus\ can be identified with only a rather
limited subset of classical \W-algebras~[\Ref{WNOS},\Ref{FHM},\Ref{FPLUS}].
However, a detailed analysis of the new cases covered by this paper, along the
same lines of Section~6 of~[\Ref{WNOS}], would be required to establish
more definite conclusions on this relevant question.

Finally, let us point out that, above all, the use of eq.~\GEND\
largely simplifies the study of the second Poisson bracket algebra of the
integrable hierarchies of~[\Ref{GEN1},\Ref{GEN2}], and it should make possible
to unveil some of their still unknown properties. For instance, in the
general case, it is still unclear whether the second Poisson bracket algebra
contains an energy-momentum tensor for some of the conformal transformations
described in Section~2, which is a necessary condition to
properly think of them as extended conformal algebras. Actually, in Sec.~7.1,
we have investigated the existence of an energy-momentum tensor for the
hierarchies constrained by the non-degeneracy condition, which
illustrates that it is indeed a very restrictive constraint. In any case, a
detailed general study of eq.~\GEND, and, in particular, of the term ${\cal
C}\{\cdot, \cdot\}$, should shed some light on this and other relevant related
questions.

\bjump\jump
\centerline{\bf Acknowledgements}

\sjump\noindent
We would like to thank Philippe Ruelle for his invaluable help at the
beginning of this work, and Laszlo Feh\'er for his very useful comments. 
We would also like to express our gratitude to Prof.~David Olive, Tim Hollowood
and to the members of the Department of Physics at Swansea for their
hospitality while part of this work was being completed. 
The research reported in this paper has been supported partially
by C.I.C.Y.T. (AEN93-0729) and D.G.I.C.Y.T. (PB93-0344).

\bjump\jump

\references
\beginref
%
%+++
\Rref{KW}{V.G.~Kac and M.~Wakimoto, {\sl Exceptional hierarchies of
soliton equations\/}, Proceedings of Symposia in Pure Mathematics {\bf
49} (1989) 191.}
%+++
\Rref{CKP}{F~Yu and Y-S~Wu, Phys. Rev. Lett. {\bf 68} (1992)
2996;\newline
Y.~Cheng, J. Math. Phys. {\bf 33} (1992) 3774;\newline
W.~Oevel and W.~Strampp, Commun. Math. Phys. {\bf 157} (1993) 51;\newline
L.~Bonora and C.S.~Xiong, Phys. Lett. {\bf B317} (1993) 329;\newline
L.A.~Dickey, {\sl On the Constrained KP Hierarchy\/}, hep-th/9407038; Lett.
Math. Phys. {\bf 35} (1995) 229.}
%
%\Rref{ZAM}{A.B.~Zamolodchikov, Theor. Math. Phys. {\bf 65} (1985) 347.}
%
\Rref{GD}{I.M.~Gel'fand and L.A.~Dikii, Funkts. Anal. Pril. {\bf 10}
(1976) 13; Funkts. Anal. Pril. {\bf 13} (1979) 13; \newline
L.A.~Dikii, {\sl Soliton equations and Hamiltonian systems\/}, Adv. Ser.
Math. Phys., Vol. 12. World Scientific, Singapore, 1991.}
\Rref{WIL}{G.W.~Wilson, Ergod. Theor. \& Dyn. Sist. {\bf 1} (1981) 361.}
\Rref{DS}{V.G.~Drinfel'd and V.V.~Sokolov, J.~Sov. Math. {\bf 30}
(1985) 1975; Soviet. Math. Dokl. {\bf 23} (1981) 457.}
%
%\Rref{FL}{S.L.~Luk'yanov and V.A.~Fateev, Int. J. Mod. Phys. {\bf A3}
%(1988) 507; Sov. Sci. Rev. A. Phys. {\bf 15} (1990) 1.} 
%
%\Rref{BAKW}{I.~Bakas, Commun. Math. Phys. {\bf 123} (1989)
%627; Phys. Lett. {\bf B213} (1988) 313;\newline
%P.~Mathieu, Phys. Lett. {\bf B208} (1988) 101.}
%
\Rref{GEN1}{M.F.~de Groot, T.J.~Hollowood and J.L. Miramontes,
Commun. Math. Phys. {\bf 145} (1992) 57.}
\Rref{GEN2}{N.J.~Burroughs, M.F.~de Groot, T.J.~Hollowood and 
J.L. Miramontes, Commun. Math. Phys. {\bf 153} (1993) 187;
Pys. Lett. {\bf B277} (1992) 89.}
\Rref{KP}{V.G.~Kac and D.H.~Peterson, {\sl Symposium on Anomalies, Geometry
and Topology}, W.A.~Bardeen and A.R.~White (eds.), Singapore, World
Scientific (1985) 276-298.}
\Rref{KACB}{V.G.~Kac, {\sl Infinite Dimensional Lie Algebras ($3^{rd}$
ed.)}, Cambridge University Press, Cambridge (1990).}
\Rref{WNOS}{C.R.~Fern\'andez Pousa, M.V.~Gallas, J.L.~Miramontes, and
J.~S\'anchez Guill\'en, Ann. Phys. (N.Y.) {\bf 243} (1995) 372.}
\Rref{TJIN}{T.~Tjin, {\sl Finite and Infinite \W-algebras and their
Applications}, PhD Thesis, Univ. of Amsterdam (1993).}
\Rref{MORZ}{N.~Jacobson, {\sl Lie Algebras}, Wiley-Interscience, New
York (1962).}
\Rref{SORB}{F.~Delduc, E.~Ragoucy, and P.~Sorba, Phys. Lett. {\bf
B279} (1992) 319.}
\Rref{NIGEL}{N.J.~Burroughs, Nonlinearity {\bf 6} (1993) 583; 
Nucl. Phys. {\bf B379} (1992) 340.}
\Rref{RMAT}{M.A.~Semenov-Tian-Shanskii, Func. Anal. Appl. {\bf 17}
(1983) 259.}
\Rref{BAK}{I.~Bakas and D.A.~Depireux, Mod. Phys. Lett. {\bf A6}
(1991) 1561, ERRATUM ibid. {\bf A6} (1191) 2351;
Int. J. Mod. Phys. {\bf A7} (1992) 1767.}
\Rref{COND}{L.~Feh\'er, L.~O'Raifeartaigh, P.~Ruelle, and I.~Tsutsui,
Phys. Lett. {\bf B283} (1992) 243.}
\Rref{POLY}{M.~Bershadsky, Commun. Math. Phys. {\bf A5} (1991) 833;
\newline
A. Polyakov, Int. J. Mod. Phys. {\bf A5} (1990) 833.}
\Rref{DMAT}{D.A.~Depireux and P.~Mathieu, Int. J. Mod. Phys. {\bf A7}
(1992) 6053.}
%
%\Rref{EXS}{P.~Mathieu and W.~Oevel, Mod. Phys. Lett. {\bf A6}
%(1991) 2397;\newline
%T.~Tjin and P.~van Driel, {\sl Coupled WZNW-Toda Models and
%Covariant KdV Hierarchies\/}, preprint ITFA-91-04;\newline
%P.~van Driel, Phys. Lett. {\bf 274B} (1991) 179; \newline
%I.~Bakas and D.A.~Depireux, {\sl Self-Duality, KdV Flows and
%W-Algebras\/}, Diff. Geom. Meth. (1991) 928; \newline
%F.~Toppan, Phys. Lett. {\bf B327} (1994) 249; \newline
%L.~Bonora and C.S.~Xiong, J.~Math. Phys. {\bf 35} (1994) 5781;
%\newline L.~Bonora, Q.P.~Liu, and C.S.~Xiong, {\sl The Integrable
%Hierarchy constructed from a pair of higher KdV Hierarchies and its
%associated W Algebra\/}, BONN-TH-94-17, hep-th 9408035.}
%
\Rref{TAU}{T.J.~Hollowood, and J.L.~Miramontes, Commun. Math. Phys. {\bf 157}
(1993) 99.}
\Rref{MAC}{I.R.~McIntosh, {\sl An Algebraic Study of Zero Curvature
Equations\/}, PhD Thesis, Dept. Math., Imperial College (London), 1988
(unpublished); J.~Math. Phys. {\bf 34} (1993) 5159.} 
\Rref{ARA}{H.~Aratyn, L.A.~Ferreira, J.F.~Gomes and A.H.~Zimerman, {\sl
Constrained KP Models as Integrable Matrix Hierarchies\/}, IFT-P/041/95,
UICHEP-TH/95-9, hep-th/9509096.}
\Rref{FHM}{L.~Feh\'er, J.~Harnad, and I.~Marshall, Commun.
Math. Phys. {\bf 154} (1993) 181; \newline
L.~Feh\'er and I.~Marshall, {\sl
Extensions of the matrix Gelfand-Dickey hierarchy from generalized
Drinfeld-Sokolov reduction\/}, SWAT-95-61, hep-th/9503217.}
\Rref{FPLUS}{L.~Feh\'er, {\sl Generalized Drinfeld-Sokolov Hierarchies and
\W-algebras}, Proceedings of the NSERC-CAP Workshop on Quantum Groups,
Integrable Models and Statistical Systems, Kingston, Canada, 1992; \newline
F.~Delduc and L.~Feh\'er, J. of Physics A--Math. and Gen. {\bf 28} (1995) 5843;
\newline L.~Feh\'er, {\sl KdV type systems and \W-algebras in the
Drinfeld-Sokolov approach\/}, talk given given at the Marseille Conference on
\W-algebra, July 1995, hep-th/9510001.}
\Rref{BOU}{P.~Bouwknegt and K.~Schoutens, Phys. Rep. {\bf 223} (1993)
183.}
\Rref{HAM}{F.A.~Bais, T.~Tjin, and P.~van Driel, Nucl. Phys. {\bf B357}
(1991) 632;\newline L.~Feh\'er, L.~O'Raifeartaigh, P.~Ruelle, I.~Tsutsui,
and A.~Wipf, Ann. Phys. (N.Y.) {\bf 213} (1992) 1.} 
\Rref{BALOG}{J.~Balog, L.~Feh\'er, P.~Forg\'acs, L.~O'Raifeartaigh, and
A.~Wipf, Ann. Phys. (N.Y.) {\bf 203} (1990) 76.}
\Rref{FREP}{L.~Feh\'er, L.~O'Raifeartaigh, P.~Ruelle,
I.~Tsutsui, and A.~Wipf, Phys. Rep. {\bf 222} (1992) 1.}
\Rref{FRENCH}{L.~Frappat, E.~Ragoucy and P.~Sorba, Commun. Math. Phys. {\bf
157} (1993) 499.}
\Rref{DSP}{L.~Feh\'er, L.~O'Raifeartaigh, P.~Ruelle, and I.~Tsutsui, 
Commun. Math. Phys. {\bf 162} (1994) 399.}  
\Rref{DIRAC}{P.A.M.~Dirac, {\sl Lectures on Quantum Mechanics\/},
Belfer Graduate School of Science, Yeshiva Univ. of New York (1964).}
%
%\Rref{LUIZ}{L.A.~Ferreira, J.F.~Gomes, A.~Schwimmer, and
%A.H.~Zimerman, Phys. Lett. {\bf B274} (1992) 65.}
%
\Rref{PRED}{R.~Abraham and J.E.~Marsden, {\sl Foundations of Classical
Dynamics ($2^{\rm nd}$ ed.)\/}, The Benjamin/Cummings Publ. Co.,
Reading, Mass. (1978);\newline  
K.~Sundermeyer, {\sl Constrained Dynamics\/}, Lecture Notes in
Physics 169, Springer-Verlag, Berlin (1982);\newline
M.~Henneaux and C.~Teitelboim, {\sl Quantization of Gauge Systems\/},
Princeton University Press, Princeton, N.J. (1992); \newline
T.~Kimura, Comm. Math. Phys. {\bf 151} (1993) 155.}
%
%\Rref{ADD}{T.J.~Hollowood, J.L.~Miramontes, and J. S\'anchez
%Guill\'en, J. Phys. A: Math. Gen. {\bf 27} (1994) 4629.}
%

\endref

\ciao

123456789012345678901234567890123456789012345678901234567890123456789012345678